\documentclass[13pt]{article}
\usepackage{amsmath}
\usepackage{amsthm}
\usepackage{amsfonts}
\usepackage{latexsym}
\usepackage[matrix,tips,graph,curve]{xy}
\usepackage{color}
\usepackage{amsfonts}
\usepackage{graphicx,amssymb}
\usepackage{multirow}
\usepackage{rotating}
\usepackage{mathrsfs}
\usepackage{natbib}
\usepackage{hyperref}
\newtheorem{thm}{Theorem}

\DeclareMathOperator*{\argmin}{argmin}

\newcommand{\bm}[1]{\mbox{\boldmath{$#1$}}}

\newcommand{\blind}{0}



\makeatletter
\@addtoreset{equation}{section}
\makeatother

\theoremstyle{plain}
\newtheorem{theorem}[equation]{Theorem}

\newtheorem{lemma}[equation]{Lemma}

\theoremstyle{definition}

\newcommand{\GG}[1]{}

\def\d/{/\mspace{-6.0mu}/}

\begin{document}
\def\spacingset#1{\renewcommand{\baselinestretch}%
{#1}\small\normalsize} \spacingset{1}
\title{\bf On buffered double autoregressive time series models}

\author{ Zhao Liu\\
Department of Economics, Duke University\thanks{\textit{E-mail: zhao.liu@duke.edu}} \\
}
\date{}

\maketitle

\if1\blind
{
  \bigskip
  \bigskip
  \bigskip
  \begin{center}
    {\LARGE\bf On buffered double autoregressive time series models}
\end{center}
  \medskip
} \fi

\bigskip
\spacingset{1.45} 
\begin{abstract}
\noindent A buffered double autoregressive (BDAR) time series model is proposed in this paper to depict the buffering phenomenon of conditional mean and conditional variance in time series. To build this model, a novel flexible regime switching mechanism is introduced to modify the classical threshold time series model by capturing the stickiness of signal. Besides, considering the inadequacy of traditional models under the lack of information, a signal retrospection is run in this model to provide a more accurate judgment. Moreover, formal proofs suggest strict stationarity and geometric ergodicity of BDAR model under several sufficient conditions. A Gaussian quasi-maximum likelihood estimation (QMLE) is employed and the asymptotic distributions of its estimators are derived. It has been demonstrated that the estimated thresholds of the BDAR model are $n$-consistent, each of which converges weakly to a functional of a two-sided compound Poisson process. The remaining parameters are $\sqrt{n}$-consistent and asymptotically normal. Furthermore, a model selection criteria and its asymptotic property have been established. Simulation studies are constructed to evaluate the finite sample performance of QMLE and model selection criteria. Finally, an empirical analysis of Hang Seng Index (HSI) using BDAR model reveals the asymmetry of investors' preference over losses and gains as well as the asymmetry of volatility structure.
\end{abstract}
{\it Some key words:} Double autoregressive model; Buffered threshold model; Geometric ergodicity.\\
{\it First version available at SSRN:} \url{https://papers.ssrn.com/sol3/papers.cfm?abstract_id=2829536}\\
{\it This version:} May 2018 

\newpage

\section{Introduction}
A buffering phenomenon refers to the variable of interest sticks in the previous state when the variable of signal stays in a certain uninformative region. Consider, for example, a decision making of treatment plan for patients with diabetes. The up-crossing of blood sugar level over a certain high level $r_U$ will intrigue an inpatient treatment, while a blood sugar level lower than $r_L$ will evoke outpatient treatment. A doctor may easily choose between two treatment plans when a patient's blood sugar level is too high ($>r_U$) or too low ($<r_L$). However, he/she may become very hesitant to make the decision when the sugar level dwells in $[r_L, r_U]$, an interval without any other explicit benchmarks or information. This chaotic interval $[r_L, r_U]$ is exactly a typical buffer zone, where a retrospection of the historical data is needed to judge the status of the variable of interest. Therefore, in this case, a wise doctor might read the patient's previous record of blood sugar level. If the blood sugar level swings around $r_U$ within the last week, the patient will be hospitalized overnight. Otherwise, the doctor will only give a prescription and require the patient to have at-home treatment.\\

\noindent Such regime switching mechanism was first formulated by \cite*{li2015}, in an effort to extend the traditional threshold AR model. Only focusing on the conditional mean function, \cite*{li2015} has not considered conditional heteroscedasticity, which is, however, an essential part in the models of financial time series. Therefore, \cite{lo2015} applied the similar idea to the conditional volatility process and proposed the buffered pure GARCH process in the study of exchange rate. \cite{zhu2015} then introduced the concept of the buffered autoregressive model with generalized autoregressive conditional heteroscedasticity (BAR-GARCH) and empirically compared its performance with AR-GARCH and Threshold-GARCH model by an analysis of exchange rate. However, the theoretical property of BAR-GARCH model remained unexplored, for which no statistical inference could be substantiated. Besides, the volatility could not be explicitly computed by observations in a direct way in BAR-GARCH model.\\

\noindent Statistical test for the existence of threshold and model selection criteria are also crucial in the construction of buffered models. A quasi-likelihood ratio for the thresholds of BAR model has been established by \cite{zhu2014}, yet similar tests for a more generalized model such as BAR-GARCH model seems not readily available. Moreover, the application of model selection criteria (AIC, $\text{AIC}_C$ and BIC) to the order selection in BAR-GARCH model lacks necessary simulation tests as well as theoretical support.\\

\noindent In this paper, we proposed a buffered double AR (BDAR) model that captures the stickiness through applying a novel regime switching mechanism established by \cite*{li2015} to double AR (DAR) model (see \cite{ling2004,ling2007}, \cite{Zhu_Ling2013}). The theoretical properties of the proposed model are thoroughly investigated. Unlike the BAR-GARCH model, the conditional variance of each regime in the BDAR model is governed directly by past observations which allows a visible dynamic behavior of conditional volatility.\\

\noindent The remainder of the paper is organized as follows. Section 2 proposes the buffered double autoregressive (BDAR) model. Formal proofs suggest strict stationarity and geometric ergodicity of BDAR model under several sufficient conditions. A Gaussian quasi-maximum likelihood estimation (QMLE) is employed to compute the estimators in this model. The strict proof reveals the asymptotic normality of the estimated coefficients governing the conditional mean and variance of each regime in the BDAR model. More importantly, it has been demonstrated that each of the estimated thresholds of the BDAR model is $n$-consistent and weakly converges to the smallest minimizer of a two-sided compound Poisson process. Our results include the buffered AR (BAR) model (\cite*{li2015}), threshold double AR (TDAR) model (\cite*{li_ling2015a}) and traditional threshold AR (TAR) model (\cite{Tong1990}) as special cases. A Bayesian-type information criteria (BIC), as well as its asymptotic property, are formally derived in section 3 for selecting orders of DAR models embedded in the BDAR model. Section 4 conducts several Monte Carlo simulation experiments to evaluate the finite sample performance of the Gaussian QMLE and proposed BIC. Finally, the model is applied to Hang Seng Index (HSI) in Section 5 to show the asymmetry of investors' preference over losses and gains as well as the leverage effect of the stock market. All technical proofs are relegated to Appendix.

\section{Buffered double autoregressive models}
Consider the following buffered double autoregressive (BDAR) model:
\begin{equation}\label{model4}
y_{t}=\left\{
\begin{array}{cc}
\phi_{10}+\sum_{i=1}^p\phi_{1i}y_{t-i} +\varepsilon_t\sqrt{\alpha_{10}+\sum_{j=1}^p\alpha_{1j}y_{t-j}^{2}} & \text{if $R_t=1$}\\
\phi_{20}+\sum_{i=1}^p\phi_{2i}y_{t-i} +\varepsilon_t\sqrt{\alpha_{20}+\sum_{j=1}^p\alpha_{2j}y_{t-j}^{2}} & \text{if $R_t=0$}
\end{array}
\right.
\end{equation}
\\
\[
R_{t}=\left(
\begin{array}{cc}
1& \text{if $y_{t-d}\leq r_L$}\\
0& \text{if $y_{t-d}> r_U$}\\
R_{t-1}& \text{otherwise}
\end{array}
\right)
\]
Where $\alpha_{k0}>0$ with $\alpha_{kj}\geq 0$ for $k=1,2$ and $j=1,\ldots,p$. And $\{\varepsilon_t\}$ are identically and independently distributed ($i.i.d.$) random variables with mean zero and variance one. The integer $d>0$ is the delay paramerter and $r_L \leq r_U$ are the boundary parameters of buffer zone.
We denote the model by BDAR(p) for simplicity. Note \eqref{model4} includes the BAR model proposed by \cite*{li2015} as a special case. When $r_L = r_U$, \eqref{model4}, the model becomes the threshold double AR model proposed by \cite{li_ling2015a}.\\

\noindent Denote $Y_t=(y_t, y_{t-1},\ldots,y_{t-p+1},R_t)^{\prime}$, $m_{it}=\phi_{1i}I(A_t)+\phi_{2i}I(A_t^c)$, where $A_t$ is the event $\{y_{t-d} \leq r_L\} \bigcup \{r_L < y_{t-d} \leq r_U, R_{t-1}=1\}$ and $A_t^c$ is its complement. After introducing notations $M_{0t}$, $M_{1t}$, $g_{1}({Y_{t-1}})$ and $g_{2}({Y_{t-1}})$ as follows, we can show that $Y_t=g_1(Y_{t-1})+\epsilon_t g_2(Y_{t-1})$ forms a Markov Chain.\\

\[M_{0t}=
\begin{bmatrix}
    \phi_{10}I(A_t)+\phi_{20}I(A_t^c)\\
    0\\
    \vdots\\
    0\\
    I(y_{t-d}\leqslant r_L)\\
\end{bmatrix}
\quad\quad
M_{1t}=
\begin{bmatrix}
    m_{1t} & m_{2t} & \dots & m_{pt} & 0\\
    1 & 0 & \dots & 0 & 0 \\
    \vdots & \vdots & \vdots & \vdots & \vdots\\
    \vdots & \vdots & \vdots & \vdots & \vdots\\
    0 & 0 & \dots &\dots & I(r_L<y_{t-d}\leqslant r_U)\\
 \end{bmatrix}
\]
\\
\begin{center}
$g_{1}(Y_{t-1})=M_{0t}+M_{1t}Y_{t-1}$\\
\end{center}

\[g_{2}(Y_{t-1})=
\begin{bmatrix}
    \sqrt{(\alpha_{10}+\sum_{j=1}^p\alpha_{1j}y_{t-j}^2)I(A_t)+(\alpha_{20}+\sum_{j=1}^p\alpha_{2j}y_{t-j}^2)I(A_t^c)}\\
    0\\
    \vdots\\
    0\\
 \end{bmatrix}
\]\\

\noindent By applying a method similar to \cite{Lee2006}, several sufficient conditions are obtained for the geometric ergodicity of $\{Y_t\}$.
\begin{thm}\label{thm1}
Suppose the distribution of $\varepsilon_t$ has a positive density $f$ over $\mathbb{R}$ and $E|\varepsilon_t|^s < \infty$ for some $s>0$. Moreover, $f$ is locally bounded away from 0 and satisfies $\sup_{x\in \mathbb{R} } \{(1+|x|)f(x)\} <\infty$,  then the multivariate process $\{Y_{t}\}$ is geometrically ergodic and hence \eqref{model4} admits a strictly stationary, geometrically ergodic solution if one of the following conditions holds:\\
(i).$\sum_{j=1}^{p} \left( \sup_{1\leq i\leq2}|\phi_{ij}|^r+\sup_{1\leq i\leq2}(\alpha_{ij}^{r/2})E|\varepsilon_t|^r\right)<1$, $r\in(0,1]$.\\
(ii).$\left(\sum_{j=1}^{p}  \sup_{1\leq i\leq2}|\phi_{ij}|\right)^r+\sum_{j=1}^{p}\sup_{1\leq i\leq2}(\alpha_{ij}^{r/2})E|\varepsilon_t|^r<1$,  $r\in(1,2]$ and $f$ symmetric.\\
(iii).$ (1+3E(\epsilon_t^2))(\sum_{j=1}^{p} \sup_{1\leq i\leq2}|\phi_{ij}|)^4+(E(\epsilon_t^4)+3E(\epsilon_t^2))(\sum_{j=1}^{p} \sup_{1\leq i\leq2}\alpha_{ij})^2<1$, $r=4$
\end{thm}
\noindent Such conditions are easy to check but may be restrictive. Obtaining more general sufficient and necessary stability conditions seems difficult and thus is left for future research.
\section{The QML estimator of BDAR(p) model}
\subsection{Estimation procedure}
This section considers the Gaussian quasi-maximum likelihood estimation of the buffered double AR model specified in \eqref{model4}.\\
Denote by $\bm{\theta}=(\bm{\lambda}',r_L, r_U, d)'$ the parameter vector of model \eqref{model4}, $\bm{\lambda}=(\bm{\phi_1'}, \bm{\alpha_1'}, \bm{\phi_2'}, \bm{\alpha_2'})$ with $\bm{\phi_i'}=(\phi_{i0},\phi_{i1},\ldots,\phi_{ip})'$ and $\bm{\alpha_i'}=(\alpha_{i0},\alpha_{i1},\ldots,\alpha_{ip})$ and \\
\begin{equation}\label{model5}
\begin{array}{cc}
\bm{u_t(\theta)}=y_t-\bm{\mu_t(\theta)},&\bm{\mu_t(\theta)}=(\bm{\phi_1'Y_{1,t-1}})R_t(r_L, r_U,d)+(\bm{\phi_2'Y_{2,t-1}})(1-R_t(r_L, r_U,d))\\
&\bm{h_t(\theta)}=(\bm{\alpha_1'X_{1,t-1}})R_t(r_L,r_U,d)+(\bm{\alpha_2'X_{2,t-1}})(1-R_t(r_L, r_U,d))
\end{array}
\end{equation}
with $\bm{Y_{1,t-1}}=\bm{Y_{2,t-1}}=\bm{Y_{t-1}}=(1,y_{t-1},\ldots,y_{t-p})'$ and $\bm{X_{1,t-1}}=\bm{X_{2,t-1}}=\bm{X_{t-1}}=(1,y_{t-1}^2,\ldots,y_{t-p}^2)'$. The conditional log-likelihood function (after multiplying -2 and omitting some constant) is defined as
\[
\bm{L_n(\theta)}=\sum_{t=1}^n \bm{l_t(\theta)},\quad \text{where} \quad  \bm{l_t(\theta)}=\text{log}\bm{h_t(\theta)}+\frac{\bm{u_{t}^2(\theta)}}{\bm{h_t(\theta)}}
\]

\noindent Let $\bm{\Lambda}$ be a compact set of $\mathbb{R}^{4p+4}$, [a b] be a predetermined interval, and $d_{max}$ be a predetermined positive integer. Assume $\bm{\lambda}\in\bm{\Lambda}$,$a \leq r_L \leq r_U \leq b$ and $d \in D=\{1,\ldots,d_{max}\}$. The true parameter vector is $\bm{\theta_0}=(\bm{\lambda_0}',r_{0L}, r_{0U}, d_0)'$ and the true regime indicator function is $R_{t0}=R_t(r_{0L}, r_{0U}, d_0)$\\

\noindent Let $n_0$=max$\{p,d_{max}\}$. For observed time series $\{y_t, -n_0 +1 \leq t\leq n\}$ generated by \eqref{model4}, the likelihood functions defined in the above depend on past observations infinitely far away, due to the novel regime switching mechanism. Hence, initial values are needed to fit the model. \\

\noindent As discussed in \cite*{li2015}, for fixed $(r_{L}, r_{U}, d)$, the first few observations of the threshold variable $y_{t-d}$, say $1\leq t\leq t_0$ may fall into the buffer zone $(r_L, r_U]$ such that the regime which $y_1, \ldots, y_{t_0}$ belong to cannot be identified. Note these $t_0$ observations come from the same regime since the threshold variables keep staying in the buffer zone. We may simply assign these $t_0$ observations to lower regime and denote the resulting regime indicator function by $\widetilde{R}_{t}(r_L,r_U,d)$. Note the value of $R_{t_{0}+1}(r_L,r_U,d)$ is known since $y_{t_{0}+1-d}$ is outside the buffer zone and it is clear $\widetilde{R}_{t}(r_L,r_U,d)=R_{t}(r_L,r_U,d)$ when $t_{0} < t \leq n$.\\

\noindent Denote by $\bm{\widetilde{l}_t(\theta)}$ and $\bm{\widetilde{L}_n(\theta)}$ the corresponding functions with $R_{t}(r_L,r_U,d)$ replaced by $\widetilde{R}_{t}(r_L,r_U,d)$. We can then define the Gaussian QMLE of the true value $\bm{\theta_0}\in \bm{\Theta}$ as $\bm{\hat{\theta}_{n}}=\arg \min_{\bm{\theta}\in \bm{\Theta}}\bm{\widetilde{L}_n(\theta)}$ .\\

\noindent One can take two steps to search $\bm{\hat{\theta}_{n}}$:\\

$\bullet$ First minimize $\bm{\hat{\theta}_{n}}$ for each fixed ($r_L,r_U,d$), i.e.
\[
\bm{\hat{\lambda}_{n}}(r_L,r_U,d)=\arg\min_{\bm{\lambda}\in \bm{\Lambda}}\bm{\widetilde{L}_n(\lambda},r_L,r_U,d)
\]

$\bullet$ Search for the estimators of ($r_L,r_U,d$) by
\[
(\hat{r}_L,\hat{r}_U,\hat{d})=\arg\min_{\substack{d\in D, \ a \leq r_L \leq r_U \leq b} } \bm{\widetilde{L}_n(\hat{\lambda}_n},(r_L,r_U,d),r_L,r_U,d)
\]
Then we obtain $\bm{\hat{\theta}_{n}}=(\bm{\hat{\lambda}'_{n}},\hat{r}_L,\hat{r}_U,\hat{d})'$.\\
\subsection{Asymptotic Results}
$\textbf{Assumption 1}$ pr$(y_t\in [a\ b])<1$, $\varepsilon_t$ is $i.i.d.$ with zero mean and unit variance, and has a positive and continuous density $f(x)$ on $\mathbb{R}$.\\

\noindent $\textbf{Assumption 2}$ The parameter space \\$\bm{\Lambda}=\left\{\bm{\lambda}\in \mathbb{R}^{4p+4}: \bm{\phi_1}\neq \bm{\phi_2}\ \text{and} \ \bm{\alpha_1} \neq \bm{\alpha_2}\right\}$ is compact. Moreover, each element in $\bm{\alpha_{1}}$ or $\bm{\alpha_{2}}$ is positive.\\ 
\begin{thm}\label{thm2}
Suppose Assumptions 1 and 2 hold and $\{y_t\}$ is strictly stationary and ergodic with $Ey_t^{2} <\infty$. Then $\bm{\hat{\theta}}_n \rightarrow  \bm{\theta}_0 \ a.s.$ as $n \rightarrow \infty$.
\end{thm}
\noindent The proof of Theorem 2 follows the standard arguments in proving strong consistency. The delay parameter d only takes integer value. By theorem 2, when the sample size n is large, $\hat{d}$ will be equal to $d_0$. As in \cite*{li2015}, we assume the true delay parameter, $d_0$, is known for the remainder of this subsection, and then it is deleted from the parameter vector $\bm{\theta}$ and corresponding functions.\\

\noindent$\textbf{Assumption 3}$ $\kappa_4\equiv E(\varepsilon_t^4)<\infty$ and $E(y_t^4)<\infty$.\\

\noindent$\textbf{Assumption 4}$ The conditional mean function $\bm{\mu_t(\theta)}$ or volatility function $\bm{h_t(\theta)}$ in \eqref{model5} are discontinuous over the buffer zone $[r_{0L},r_{0U}]$, i.e. there exist $p-1$ constants $z_{p-1}$,$\ldots$,$z_{p-d+1}$,$z_{p-d-1}$,$\ldots$,$z_{0}$ such that
\[
\{(\bm{\phi_{10}}-\bm{\phi_{20}})'\bm{z}\}^{2}+ \{(\bm{\alpha_{10}}-\bm{\alpha_{20}})^{\prime}\bm{Z}\}^{2}>0
\]
for all $z_{p-d}\in [r_{0L},r_{0U}]$ where $\bm{z}=(1,z_{p-1},\ldots,z_0)'$ and $\bm{Z}=(1,z_{p-1}^2,\ldots,z_0^2)'$.  Without loss of generality, here we assume that $d\leq p$.\\

\noindent Let $Y_t=(y_t,\ldots,y_{t-p+1},R_{t})'$,then $\{Y_t\}$ is a Markov chain as shown in previous section. Denote its m-step transition probability function by $P^m(x,A)$, where $x \in \mathbb{R}^{p} \times \{0,1\}$, $A\in \mathcal{B}_p \times \mathcal{U}$, where $\mathcal{B}_p$ is the class of Borel sets of $\mathbb{R}^{p}$ and $\mathcal{U}=\{{\O},\{0\},\{1\},\{0,1\}\}$\\

\noindent$\textbf{Assumption 5}$ The time series $\{Y_t\}$ admits a unique invariant measure $\pi(\cdot)$, such that there exist $K>0$ and $0 \leq \rho <1$ for any $x\in \mathbb{R}^p \times \{0,1\}$ and any m, $\|P^m(x,\cdot)-\pi(\cdot)\|_v \leq K(1+\|x\|^2)\rho^m$, where $\|\cdot\|_v$ and $\|\cdot\|$are respectively the total variation norm and Euclidean norm.\\

\noindent Under Assumption 5, $\{Y_t\}$ is said to be V-uniformly ergodic with $V(x)=K(1+\|x\|^2)$, a condition stronger than geometric ergodicity.\\ 
\begin{thm}\label{thm3}
Suppose Assumptions 1 to 5 hold and $\bm{\theta_0}$ is an interior point of $\bm{\Theta}$. Then:\\
(i). $n(\hat{r}_L-r_{0L})=O_p(1)$ and $n(\hat{r}_U-r_{0U})=O_p(1)$\\
(ii). $\sqrt{n}\sup_{\substack{n(|r_U-r_{0U}|+|r_L-r_{0L}|)\leq B}}\|\bm{\hat{\lambda}_n}(r_L,r_U)-\bm{\hat{\lambda}_n}(r_{0L},r_{0U})\|=o_p(1)$ for any fixed constant $0 <B < \infty$, where $\bm{\hat{\lambda}_n}(r_L,r_U)$ is the QMLE given $r_L,r_U$ are known. Furthermore, it follows that:\\
$\sqrt{n}(\bm{\hat{\lambda}_n}-\bm{\lambda_0})=\sqrt{n}(\bm{\hat{\lambda}_n}(r_{0L},r_{0U})-\bm{\lambda_0})+o_p(1) \Rightarrow N(\bm{0,\Omega^{-1}\Sigma\Omega^{-1}})$, as $n \rightarrow \infty$.\\
Where $\Omega=\text{diag}(\bm{A_1,0.5B_1,A_2,0.5B_2}),\bm{\Sigma=\text{diag}(\Sigma_1,\Sigma_2)}$ with\\
\[
\bm{\Sigma_i=\left(
\begin{array}{cc}
A_i& \frac{\kappa_3}{2}D_i\\
\frac{\kappa_3}{2}D_i'& \frac{\kappa_4-1}{4}B_i
\end{array}
\right)},
\]
where $\kappa_3=E(\varepsilon_1^3)$\\
\[
\bm{A_i}=E\left\{\bm{\frac{Y_{i,t-1}Y_{i,t-1}'}{\alpha_{i0}'X_{i,t-1}}}g_{i}(r_{0L},r_{0U})\right \}\\
\bm{B_i}=E\left\{\bm{\frac{X_{i,t-1}X_{i,t-1}'}{(\alpha_{i0}'X_{i,t-1})^2}}g_{i}(r_{0L},r_{0U})\right\}\\
\]
\[
\bm{D_i}=E\left\{\bm{\frac{Y_{i,t-1}X_{i,t-1}'}{(\alpha_{i0}'X_{i,t-1})^{3/2}}}g_{i}(r_{0L},r_{0U})\right \}
\]\\
with $g_{1}(r_{0L},r_{0U})=R_t(r_{0L},r_{0U})$ and $g_{2}(r_{0L},r_{0U})=1-R_t(r_{0L},r_{0U})$
\end{thm}
\noindent To study the limiting distribution of $\hat{r}_L$ and $\hat{r}_U$, denote the $\xi_{1t}=\sum\limits_{j=0}^\infty \zeta_{1, t+j} H_{t+j, j}$ and $\xi_{2t}=\sum\limits_{j=0}^\infty \zeta_{2, t+j} H_{t+j, j}$, where\\
$\begin{aligned}
\zeta_{1t}&=\log\frac{\bm{\alpha_{10}^\prime X_{t-1}}}{\bm{\alpha_{20}^\prime X_{t-1}}}+\frac{\{\bm{(\phi_{10}-\phi_{20})^\prime Y_{t-1}}-\epsilon_t \bm{\sqrt{\alpha_{20}^\prime X_{t-1}}}\}^2}{\bm{\alpha_{10}^\prime X_{t-1}}}-\epsilon_t^2\\
\zeta_{2t}&=\log\frac{\bm{\alpha_{20}^\prime X_{t-1}}}{\bm{\alpha_{10}^\prime X_{t-1}}}+\frac{\{\bm{(\phi_{10}-\phi_{20})^\prime Y_{t-1}}+\epsilon_t \sqrt{\bm{\alpha_{10}^\prime X_{t-1}}}\}^2}{\bm{\alpha_{20}^\prime X_{t-1}}}-\epsilon_t^2
\end{aligned}$\\
$H_{t,j}=\prod\limits_{l=1}^j 1(r_{0L}<y_{t-d+1-l}\leqslant r_{0U})$ with convention $\prod\limits_{l=1}^0=1$\\

\noindent Similar to 	\cite*{li2015}, for $i=1,2$, denote $F_{i,L}(\cdot | r)$ and $F_{i,U}(\cdot | r)$ be the conditional distribution functions of $\xi_{it}\{1-R_{t-1}(r_0)\}$ and $\xi_{it}R_{t-1}(r_0)$ given $y_{t-d}=r$ respectively, where $r_0=(r_{0L},r_{0U})'$. Denote $\pi(\cdot)$ the density function of $y_t$. Then define two independent one-dimensional two-sided compound Poisson process:\par
       \begin{center}
         $\varphi_L(t)=1(t \geqslant 0)\sum\limits_{k=1}^{N_1^{(L)}(|t|)} \chi_k^{(1,L)}+ 1(t < 0)\sum\limits_{k=1}^{N_2^{(L)}(|t|)} \chi_k^{(2,L)}$\\
         $\varphi_U(t)=1(t \geqslant 0)\sum\limits_{k=1}^{N_1^{(U)}(|t|)} \chi_k^{(1,U)}+ 1(t < 0)\sum\limits_{k=1}^{N_2^{(U)}(|t|)} \chi_k^{(2,U)}$\\
       \end{center}
where $\chi_k^{(i,L)}$ and $\chi_k^{(i,U)}$ have conditional distribution functions  $F_{i,L}(\cdot | r_{0L})$ and $F_{i,U}(\cdot | r_{0U})$ respectively, $i=1,2$. $\{N_1^{(L)}(t), t\geqslant 0\},\{N_2^{(L)}(t), t\geqslant 0\}, \{N_1^{(U)}(t), t\geqslant 0\}, \{N_2^{(U)}(t), t\geqslant 0\}$ are four independent Poisson Processes with $N_1^{(L)}(0)=N_2^{(L)}(0)=N_1^{(U)}(0)=N_2^{(U)}(0)=0$. $N_1^{(L)}(\cdot), N_2^{(L)}(\cdot)$ have jump rate $\pi(r_{0L})$, while $N_1^{(U)}(\cdot), N_2^{(U)}(\cdot)$ have jump rate $\pi(r_{0U})$. Moreover, $N_1^{(L)}(\cdot)$ and $N_1^{(U)}(\cdot)$ are right continuous. $N_2^{(L)}(\cdot)$ and $N_2^{(U)}(\cdot)$ are left continuous. \\

\noindent It is implied by Assumption 4 that $E[\chi_k^{(1,j)}]$ and $E[\chi_k^{(2,j)}]>0$ for $j=L$ and $U$. Then $\varphi_j(t) \to \infty$ a.s. as $|t|\to \infty$. Then define $\varphi(z)=\varphi_L(z_L)+\varphi_U(z_U)$, where $z=(z_L, z_U)\in R^2$. As in \cite*{li2015} and \citet*{li2015b}, there exists a unique random square $[M^{(L)}_-,M^{(L)}_+) \times [M^{(U)}_-,M^{(U)}_+)$ on which $\varphi(z)$ attains the global minimum, where $[M^{(j)}_-,M^{(j)}_+)=\argmin_{t\in R} \varphi_j(t)$. \\

\noindent The following theorem states that $n(\hat{r}_L-r_{0L})$ and $n(\hat{r}_U-r_{0U})$ converges weakly to a functional of the compound Poisson Process $\varphi(z)$.\\

\begin{thm}\label{thm4}
If Assumption 1 to 5 hold, then $n(\hat{r}_L-r_{0L}) \to M^{(L)}_-$ and $n(\hat{r}_U-r_{0U}) \to M^{(U)}_-$. Furthermore, $n(\hat{r}_L-r_{0L}), n(\hat{r}_U-r_{0U})$ and $\sqrt{n}(\bm{\hat{\lambda}}_n-\bm{\lambda}_0)$ are asymptotically independent. \\
\end{thm}

\subsection{Model Selection}
We consider the Bayesian Information Criteria (BIC) for model selection, since BIC outperforms AIC for buffered type models, see \cite*{li2015}.\\

\noindent Write for simplicity $\tilde {R}_t$ as the QMLE estimator of $\tilde {R}_t(\hat{r}_L,\hat{r}_U,\hat{d})$. $n_1=\sum\limits_{t=1}^{n}\tilde {R}_t$ and $n_2=n-n_1$ Then define\\

$\begin{aligned}
\text{BIC(p)}&=\sum\limits_{t=1}^n \bm{l_t(\hat{\theta})}+(2p+2)\log n_1+(2p+2)\log n_2
\end{aligned}$\\

\noindent where $\bm{l_t(\theta)}$ is defined in Section 3.1.  Let $p_0$ be the true order of the model and $p_{\text{max}}$ be a predetermined large order. $\hat{p}_n=\argmin_{0\leq p\leq p_{\text{max}}}\text{BIC(p)}$ Then,

\begin{thm}\label{thm5}
If Assumption 1 to 5 hold, then $\text{Pr}(\hat{p}_n=p_0)\rightarrow 1$ as $n \rightarrow \infty$.\\
\end{thm}
\section{Simulation studies}
We first examine the finite sample performance of QMLE. We use sample size $n=400, 800$, each with replications 500 for the following model\\
\begin{equation}\label{sim-eq1}
y_t=\left\{
\begin{array}{rc}
-0.1+0.2y_{t-1}+0.1y_{t-2}+\varepsilon_t\sqrt{0.1+0.3y_{t-1}^2+0.05y_{t-2}^2} & \hspace{10mm} R_{t}=1\\
0.1-0.2y_{t-1}+0.3y_{t-2}+\varepsilon_t\sqrt{0.05+0.2y_{t-1}^2+0.1y_{t-2}^2} & \hspace{10mm} R_{t}=0
\end{array}
\right.
\end{equation}
with the regime indicator
\[
R_{t}=\left(
\begin{array}{cc}
1& \text{if $y_{t-4}\leq -0.1$}\\
0& \text{if $y_{t-4}> 0.15$}\\
R_{t-1}& \text{otherwise}
\end{array}
\right)
\]
where $\{\varepsilon_t\}$ are independently and identically distributed with standard normality. The estimation procedure discussed in Section 3.1 is employed. The range of the boundary parameters $r_L$ and $r_U$ is set from 10th percentile to 90th percentile of each sample. And the maximum delay parameter $d_{max}$ is set to be 6. For each sample size $n=400, 800$, $d$ can be correctly identified for all 500 replications respectively. The bias, the empirical standard deviation (ESD) and the asymptotic standard deviation (ASD) are listed in Table 1. We can observe the bias and the empirical standard deviation decrease as the sample size increases and all the ESDs are close to corresponding ASDs, which are consistent with the asymptotic results in Theorem 2 and Theorem 3.  Note the ASD of boundary parameters $r_L$ and $r_U$ cannot be estimated and only the ESD are included in Table 1. We observe the ESD of $r_L$ and $r_U$ are approximately reduced by one half when the sample size doubles, which is evidence supporting the super-consistency of boundary parameters. For each sample size, the histograms of $n(\hat{r}_L-r_{0L})$ and $n(\hat{r}_U-r_{0U})$ are displayed in Figure 1, whose shapes are close to the ones reported in \cite{Li2013} and \cite*{li2015} \\

\noindent We now evaluate the finite sample performance of model selection criteria (BIC) in Section 3.3. The data generating process is the same as ($\ref{sim-eq1}$). The sample size is 800, and there are 150 replications. The maximum order $p_{max}$ for each regime's conditional mean and conditional variance is set to be 6. And the maximum delay parameter $d_{max}$ is still set to be 6. As a result, the BIC correctly identifies all true orders at a rate of 100$\%$.
\section{Empirical Analysis}
This section considers the weekly closing prices of Hang Seng Index over the period January 2000 to December 2007. We focus on the log return in percentage $y_t=100(\log P_t-\log P_{t-1})$, where $P_t$ is the weekly closing price at time t. There are 417 observations of $\{y_t\}$ in total. The plot of $\{P_t\}$ and $\{y_t\}$ are displayed in Figure 2. \cite{li_ling2015a} have studied this series and they conclude the existence of nonlinearity, ARCH effect and threshold effect through Tsay's test (\cite{Tsay1986}), Mcleod-Li's test (\cite{Mcleod1983}) and their score-based test respectively. The two-regime threshold double autoregressive model (TDAR) is employed to fit the data in \cite{li_ling2015a}. Since TDAR model can be considered as a special case of BDAR model, it motivates us to consider the proposed BDAR model to fit the percentage log return series. \\

\noindent The range of boundary parameters $r_L$ and $r_U$ is from the 10th to the 90th empirical percentiles of observations. The maximum delay parameter $d_{max}$ is set to be 6 and the maximum order $p_{max}$ for each regime's conditional mean and conditional variance is set to be 5. Based on BIC, we have the following fitted model:
\begin{equation}\label{empirical-eq1}
y_t=\left\{
\begin{array}{lc}
0.3937_{0.2621}+0.0385_{0.1318}y_{t-1}+0.2093_{0.0894}y_{t-2}+\varepsilon_t\sigma_t & \hspace{10mm} R_{t}=1\\
-0.5992_{0.2665}+0.2354_{0.0970}y_{t-1}+\varepsilon_t\sigma_t & \hspace{10mm} R_{t}=0
\end{array}
\right.
\end{equation}
where\\
\[
\sigma_t^{2}=\left\{
\begin{array}{lc}
5.3991_{0.9225}+0.5432_{0.1583}y_{t-1}^{2}+0.1787_{0.1161}y_{t-2}^2 & \hspace{10mm} R_{t}=1\\
3.4473_{0.7086}+0.0263_{0.0599}y_{t-1}^2+0.0678_{0.0578}y_{t-2}^2+0.1416_{0.0631}y_{t-3}^2 & \hspace{10mm} R_{t}=0
\end{array}
\right.
\]
with the regime indicator\\
\[
R_{t}=\left\{
\begin{array}{cl}
1& \text{if $y_{t-1}\leq -0.2048$}\\
0& \text{if $y_{t-1}> 0.8770$}\\
R_{t-1}& \text{otherwise}
\end{array}
\right.
\]
where the subscripts of parameter estimates are their associated standard errors. Figure 3 displays respectively ACFs of residuals and squared residuals, and they slightly go beyond the 95$\%$ confidence bands only at a few lags, which partially suggests the adequacy of the fitted model. To further investigate the fit-adequacy, the Ljung-Box test statistic $Q_m$ and the McLeod-Li test statistic $\widetilde{Q}_m$ are employed. Generally, m takes value 6 and 12, see \cite{Tse2002} for a discussion on the choice of m. The p-values for of $Q_6$, $Q_{12}$, $\widetilde{Q}_6$ and $\widetilde{Q}_{12}$ are 0.6590, 0.3666, 0.8200 and 0.4013 respectively, which further suggests the fit-adequacy at 5$\%$ significance level.\\

\noindent The asymmetry of the buffered zone around 0 might be interpreted as market participants are more sensitive to losses and gains. Based on the fitted model, the investors require the return of previous week to go above 0.8770$\%$ to confirm a good time, approximately three times larger in absolute value than the signal of a bad time. When the previous week's return lies in the buffer zone [-0.2048$\%$ 0.8770$\%$], investors are not sure about the market condition, so they trace back to learn $R_{t-1}$, which may depend on returns several periods ago. Therefore, the buffered double autoregressive model will further rely on the information contained in the dynamic momentum structure, when the past week's return alone is not quite informative on the market conditions. The BDAR model provides a more realistic characterization on the regime switching mechanism than the traditional two-regime threshold type model. The sudden switch of the probabilistic structure as in the traditional threshold time series models can be delayed by the incorporation of the buffer zone.\\

\noindent Another feature of the fitted BDAR model is the asymmetry of volatility. The coefficients in the lower regime's conditional variance is much larger than their counterparts in the upper regime. This asymmetry can be interpenetrated as the stock market is more volatile during bad times.
\newpage
\section{Appendix}
\subsection{Proof of Theorem 1}
 As in the main part of the paper, define
 \[Y_t=
\begin{bmatrix}
    y_t\\
    y_{t-1}\\
    \vdots\\
    y_{t-p+1}\\
    R_t\\
\end{bmatrix}
\]\\

\[M_{0t}=
\begin{bmatrix}
    \phi_{10}I(A_t)+\phi_{20}I(A_t^c)\\
    0\\
    \vdots\\
    0\\
    I(y_{t-d}\leqslant r_L)\\
\end{bmatrix}
\quad\quad
M_{1t}=
\begin{bmatrix}
    m_{1t} & m_{2t} & \dots & m_{pt} & 0\\
    1 & 0 & \dots & 0 & 0 \\
    \vdots & \vdots & \vdots & \vdots & \vdots\\
    \vdots & \vdots & \vdots & \vdots & \vdots\\
    0 & 0 & \dots &\dots & I(r_L<y_{t-d}\leqslant r_U)\\
 \end{bmatrix}
\]
\\
\begin{center}
$g_{1}(Y_{t-1})=M_{0t}+M_{1t}Y_{t-1}$\\
\end{center}

\[g_{2}(Y_{t-1})=
\begin{bmatrix}
    \sqrt{(\alpha_{10}+\sum_{j=1}^p\alpha_{1j}y_{t-j}^2)I(A_t)+(\alpha_{20}+\sum_{j=1}^p\alpha_{2j}y_{t-j}^2)I(A_t^c)}\\
    0\\
    \vdots\\
    0\\
 \end{bmatrix}
\]\\

\begin{center}
$Y_t=g_1(Y_{t-1})+\epsilon_t g_2(Y_{t-1}) \quad \text{   forms a Markov Chain}$\\
\end{center}

Let $\mathcal{B}^p$ be the class of Borel sets of $\mathcal{R}^p$, $\mu_{p+1}$ be the Lebesgue measure on $\mathcal{R}^{p+1}$ and $\mathcal{U}=\{\emptyset, \{0\}, \{1\}, \{0,1\}\}$\\

We denote its state space by $(\mathcal{R}^p \times \{0,1\}, \mathcal{B}^p \times \mathcal{U}, \mu_{p+1})$ and sets its transition probability function as:
\begin{center}
$P(x, A)=\int_{A_1}f_{\epsilon}(y)dy, \quad x\in \mathcal{R}^p \times \{0,1\} \text{ and }A\in \mathcal{B}^p \times \mathcal{U}$\\
\end{center}
Where $A_1=\{\epsilon: \epsilon g_2(x)+g_1(x) \in A \}$ and $f_{\epsilon}(\cdot)$ is the density of $\epsilon_t$.\\

By a method similar to Lemma 1 and 2 of \cite{Lu1998}, it can be shown that the chain is $\mu_{p+1}$-irreducible and aperiodic, and non-null compact sets are small sets.\\

Before we proceed, we need to state one lemma and one theorem. The lemma is proved in \cite{Lee2006} and the theorem comes from \cite{Tweedie1983}:\\
\begin{lemma}
\label{lemma 1}
\emph{(Lemma 2.2 in \cite{Lee2006})}
Let $V(z)=\sum_{i=1}^n \gamma_i |z_i|^r, z=(z_1, z_2, \dots, z_n), n\in Z^{+}, r>0$. If $\sum_{i=1}^n \xi_i <1$ with $\xi_i\geqslant 0$, we may choose $\gamma_i>0, i=1, \dots, n$ so that for some positive constant $\rho<1, \gamma_1(\sum_{i=1}^{n}\xi_i|z_i|^r)+\sum_{i=2}^{n}\gamma_i|z_{i-1}|^r \leqslant \rho V(z)$.\\
\end{lemma}

\begin{theorem}
\emph{(Theorem 4 in \cite{Tweedie1983})}
\label{M and T}
Suppose that the Markov process $\{Y_t: t\geqslant 0\}$ is aperiodic $\mu$-irreducible and $\mathcal{B}$ is a small set. Suppose there are constants $\rho<1, \epsilon>0$ and a measurable function $V\geqslant 1$ such that\\
\begin{center}
 $E[V(Y_t)|Y_{t-1}=z]\leqslant \rho V(z)-\epsilon, z\in \mathcal{B}^c$\\
 \quad\\
 $\sup_{z\in\mathcal{B}}E[V(Y_t)|Y_{t-1}=z]<\infty$\\
\end{center}
then the Markov process $Y_t$ is geometrically ergodic.\\
\end{theorem}
\quad \\
We continue the proof. Let $b_j=\underset{1\leqslant i\leqslant 2}{max}|\phi_{ij}|, 0\leqslant j\leqslant p, \quad a_j=\underset{1\leqslant i\leqslant 2}{max} \alpha_{ij}, 0\leqslant j\leqslant p$. Define a test function $V$ by:\\
\begin{center}
$V(Y_t)=\sum\limits_{i=1}^p \gamma_i |y_{t+1-i}|^r+1 $\\
\end{center}
$|y_t| \leqslant b_0+\sum\limits_{j=1}^p b_j|y_{t-j}|+|\epsilon_t|(a_0+\sum\limits_{j=1}^p a_j y_{t-j}^2)^{\frac{1}{2}}$\\

Define:
\begin{align*}
           &s_1(y_{t-1}, \dots, y_{t-p})=b_0+\sum\limits_{j=1}^p|y_{t-j}|b_j\\
           & s_2(y_{t-1}, \dots, y_{t-p})=[a_0+\sum\limits_{j=1}^p y_{t-j}^2 a_j]^{\frac{1}{2}}\\
\end{align*}

 Similar to \cite{Lee2006}, we have:\\

\begin{align*}
&s_1^r \leqslant \left\{
\begin{aligned}
&b_0^r+\sum\limits_{j=1}^p b_j^r|y_{t-j}|^r, & & 0<r\leqslant 1\\
&(1+\epsilon)^r(\sum\limits_{j=1}^p b_j)^{r-1}(\sum\limits_{j=1}^p b_j |y_{t-j}|^r), & & 1<r\leqslant 2, \forall \epsilon>0, \exists M(\epsilon) s.t. ||Y_{t-1}||>M(\epsilon)\\
\end{aligned}
\right.
\\
&s_2^r\leqslant a_o^{\frac{r}{2}}+\sum\limits_{j=1}^p a_j^{\frac{r}{2}}|y_{t-j}|^r, \quad 0<r\leqslant 2\\
\end{align*}

Moreover, when $0<r\leqslant 1$, by basic inequality,  we have  $E[|y_t|^r|Y_{t-1}] \leqslant s_1^r+s_2^rE|\epsilon_t|^r$\\

When $1<r\leqslant 2$, note $\epsilon_t$ is symmetric by assumption, by the following inequality:
 \begin{align*}
 &(1+x)^r+(1-x)^r\leqslant 2(|x|^r+1), -1\leq x \leq 1
 \end{align*}

we can show that
 \[E[|s_1+s_2\epsilon_t|^r|Y_{t-1}]=\frac{1}{2}E[|s_1+s_2\epsilon_t|^r+|s_1-s_2\epsilon_t|^r|Y_{t-1}]\leqslant s_1^r+s_2^r E|\epsilon_t|^r\]

$\begin{aligned}
\text{Therefore, }&E[V(Y_t)|Y_{t-1}]\\
&\leqslant \gamma_1(s_1^r+s_2^r E|\epsilon_t|^r)+\sum\limits_{i=2}^p \gamma_i|y_{t+1-i}|^r+1\\
&\leqslant \gamma_1 \sum\limits_{i=1}^p\xi_i |y_{t-i}|^r+\sum\limits_{i=2}^p \gamma_i |y_{t+1-i}|^r+1+C_1,\\ &\quad \text{   where } C_j \text{ is a generic notation for positive constants}\\
 \end{aligned}$
\\
\quad\\

$ \sum\limits_{i=1}^p \xi_i=\left\{
\begin{aligned}
&\sum\limits_{j=1}^p b_j^r+\sum\limits_{j=1}^p a_j^{\frac{r}{2}}E|\epsilon_t|^r, & & 0<r\leqslant 1 \\
&(1+\epsilon)^r (\sum\limits_{j=1}^p b_j)^r+\sum\limits_{j=1}^p a_j^{\frac{r}{2}}E|\epsilon_t|^r, & &1<r\leqslant 2, \forall \epsilon>0, \exists M(\epsilon), s.t. ||Y_{t-1}||>M(\epsilon)\\
\end{aligned}
\right.
$\\
\quad\\

\noindent By Lemma \ref{lemma 1}, we can set $\sum\limits_{i=1}^p \xi_i<1$ and find $\gamma_i>0, i=1,\dots, p$ such that for $||z||>M,\exists $ some constants $\rho<1, E[V(Y_t)|Y_{t-1}=z] \leqslant \rho V(z)+C_2$. Thus, inequalities in Theorem \ref{M and T} hold with some $\epsilon>0$ and compact set $\mathcal{B}=\{||z||\leqslant M\}$ for sufficiently large $M<\infty$, since $V(z)$ increases as $||z||$ increases. Hence, we obtain the geometric ergodicity and hence the strict stationarity.\\

\noindent To prove the third part of the theorem, define $V(z)=\sum\limits_{i=1}^p \gamma_i z_i^4+1$ from $E(\epsilon_t)=E(\epsilon_t^3)=0$ for $||z||>M$.\\

$\begin{aligned}
&E[(s_1+s_2\epsilon_t)^4|Y_{t-1}] \leqslant [(1+3E(\epsilon_t^2))s_1^4+(E\epsilon_t^4+3E\epsilon_t^2)s_2^4]\\
&s_1^4 \leqslant (1+\epsilon)^4(\sum\limits_{j=1}^p b_j y_{t-j}^4)(\sum\limits_{j=1}^p b_j)^3\\
&s_2^4 \leqslant (1+\epsilon)^2(\sum\limits_{j=1}^p a_j y_{t-j}^4)(\sum\limits_{j=1}^p a_j)\\
\end{aligned}$
\\

Therefore for $||z||>M$\\

$\begin{aligned}
E[V(Y_t)|Y_{t-1}=z] &\leqslant \gamma_1\{[1+3E(\epsilon_t^2)]s_1^4+[E(\epsilon_t^4)+3E(\epsilon_t^2)]s_2^4\}+\sum\limits_{i=2}^p \gamma_i |y_{t+1-i}|^4+1\\
  &\leqslant \gamma_1\sum\limits_{i=1}^p \xi_i|y_{t-i}|^4+\sum\limits_{i=2}^p \gamma_i |y_{t+1-i}|^4+1\\
\end{aligned}$

where $ \sum\limits_{i=1}^p \xi_i=[1+3E(\epsilon_t^2)](1+\epsilon)^4(\sum\limits_{j=1}^p b_j)^{4}+[E(\epsilon_t^4)+3E(\epsilon_t^2)](1+\epsilon)^2(\sum\limits_{j=1}^p a_j)^2$\\
\noindent  By Lemma \ref{lemma 1}, we can set $\sum\limits_{i=1}^p \xi_i<1$ and obtain the geometric ergodicity as we do in proving the first two parts of the theorem.


\subsection{Proof of Theorem 2}

Before we prove Theorem 2, we introduce some notations and a preliminary result.\\

\noindent Denote the parameter space as $\bm{\Theta=\land\times[a,b] \times [a,b] \times D$, where $\land=\land_1 \times \land_2}$.\\
Parametre vector $\bm{\theta=(\phi_1^\prime, \alpha_1^\prime, \phi_2^\prime, \alpha_2^\prime, r_L, r_U, d)^\prime=(\lambda_1^\prime, \lambda_2^\prime, r_L, r_U, d)^\prime}$, where\\
  \[\bm{\lambda_1}=
\begin{bmatrix}
    \bm{\phi_1}\\
    \bm{\alpha_1}\\
\end{bmatrix}
\quad\quad
\bm{\lambda_2}=
  \begin{bmatrix}
   \bm{\phi_2}\\
   \bm{\alpha_2}\\
  \end{bmatrix}
\]

\noindent Recall in the main body of the paper, we define\\
$\begin{aligned}
 \bm{l_t(\theta)} &=\log{\bm{h_t}(\theta)}+\frac{\bm{u_t^2(\theta)}}{\bm{h_t(\theta)}}\\
     &=R_t\{\log \bm{h_{1t}(\theta)}+\frac{[y_t-\bm{\mu_{1t}(\theta)]}^2}{\bm{h_{1t}(\theta)}}\}+(1-R_t)\{{\log \bm{h_{2t}(\theta)}+\frac{[y_t-\bm{\mu_{2t}(\theta)]}^2}{\bm{h_{2t}(\theta)}}}\}\\
\end{aligned}$
\quad\\

\noindent  where for i=1,2

$\begin{aligned}
    \bm{\mu_{it}(\theta)}&=\phi_{i0}+\phi_{i1}y_{t-1} + \dots + \phi_{ip}y_{t-p}\\
    \bm{h_{it}(\theta)}&=\alpha_{i0}+\alpha_{i1}y_{t-1}^2 + \dots + \alpha_{ip}y_{t-p}^2\\
\end{aligned}$\\

\noindent We further denote it as $\bm{l_t(\theta)}=R_t \bm{l_{1t}(\theta)}+(1-R_t)\bm{l_{2t}(\theta)}$, with \\

$\begin{aligned}
     \bm{l_{1t}(\theta)}&=\log \bm{h_{1t}(\theta)}+\frac{[y_t-\bm{\mu_{1t}(\theta)]}^2}{\bm{h_{1t}(\theta)}}\\
     \bm{l_{2t}(\theta)}&={\log \bm{h_{2t}(\theta)}+\frac{[y_t-\bm{\mu_{2t}(\theta)]}^2}{\bm{h_{2t}(\theta)}}}\\
\end{aligned}$\\

\noindent If Assumption 2 and conditions in Theorem 2 hold,  Lemma B.2 in \cite{ling2007} implies that\\
\begin{equation}
\label{assumption 2}
E[\underset{\theta\in \Theta}{\sup} |\bm{l_{it}(\theta)}|]<\infty \quad \text{i=1,2 and} \quad E[\underset{\theta\in \Theta}{\sup} |\bm{l_t(\theta)}|]<\infty\\
\end{equation}

\noindent This preliminary result will be used to invoke dominance convergence theorem in the following. We are ready to prove Theorem 2 now. \\

\noindent Proof of Theorem 2\\

\noindent Following the method in Huber(1967), it is sufficient for us to verify the following three claims:\\
\begin{itemize}
  \item $S1$: $\underset{\theta\in \Theta}{\sup} \frac{1}{n}|\bm{\tilde{L}_n(\theta)-L_n(\theta)}|\xrightarrow{a.s.} 0$, where $\bm{\tilde{L}_n(\theta)}$ is the modified likelihood function defined in the paper.\\
  \item $S2$: $E[\bm{l_t(\theta)}]$ is uniquely minimized at $\bm{\theta=\theta_0=(\phi_{10}^\prime, \alpha_{10}^\prime,\phi_{20}^\prime, \alpha_{20}^\prime,r_{0L}, r_{0U}, d_0)^\prime}$\\
  \item $S3$: $E[\underset{\bm{\theta^*}\in U_{\eta}(\bm{\theta})}{\sup}|\bm{l_t(\theta)-l_t(\theta^*)}|]\xrightarrow{a.s.} 0$ as $\eta \to 0$, where $U_{\eta}(\bm{\theta})=\{\bm{\theta^*\in \Theta, ||\theta^*-\theta||}<\eta\}$\\
\\
  \end{itemize}

\noindent Let us first show S1:\\

  $\begin{aligned}
  \frac{1}{n}|\bm{\tilde{L}_n(\theta)-L_n(\theta)}|&=\frac{1}{n}|\sum\limits_{t=1}^{k_n}(\bm{\tilde{l}_t(\theta)-l_t(\theta)})|\\
     &=\{(1-R_1) \prod\limits_{t=1}^{k_n} 1(r_L<y_{t-d}\leqslant r_U)\}\{\frac{1}{n}\sum\limits_{t=1}^{k_n}[\bm{l_{1t}(\theta)-l_{2t}(\theta)}]\}\\
     &\leqslant \prod\limits_{t=1}^{k_n} 1(a \leqslant y_{t-d} \leqslant b )\{\frac{1}{n} \sum\limits_{t=1}^{k_n}[\bm{l_{1t}(\theta)-l_{2t}(\theta)}]\}\\
  \end{aligned}$\\

\noindent  If $\frac{k_n}{n} \to 0$, then clearly $\underset{\tiny{\bm{\theta \in \Theta}}}{\sup} \frac{1}{n}|\bm{\tilde{L}_n(\theta)-L_n(\theta)}|\to 0$\\

\noindent  If $\frac{k_n}{n} \to$ a positive number (constant), then by ergodic theorem, Assumption 1 in the paper and (\ref{assumption 2}), we have:\\
 {\fontsize{9bp}{\baselineskip}\begin{center}
  $\underset{\theta\in \Theta}{\sup}\frac{1}{k_n}\sum\limits_{t=1}^{k_n}[\bm{l_{1t}(\theta)-l_{2t}(\theta)}] \xrightarrow{a.s.} E[\underset{\theta\in \Theta}{\sup}[\bm{l_{1t}(\theta)-l_{2t}(\theta)}] \leqslant E[\underset{\theta\in \Theta}{\sup}[\bm{l_{1t}(\theta)}]+E[\underset{\theta\in \Theta}{\sup} \bm{l_{2t}(\theta)}]<\infty$\\
  \end{center}}

\noindent Moreover, by Assumption 1,
\begin{center}
$\frac{1}{k_n}\sum\limits_{t=1}^{k_n} 1(y_{t-d}<a \text{ or } y_{t-d}>b) \to Pr(y_{t-d}<a \text{ or } y_{t-d}>b)>0$\\
\end{center}

$Pr(\underset{k_n \to \infty}{\lim}\prod\limits_{t=1}^{k_n} 1(a\leqslant y_{t-d}\leqslant b)=1)=Pr(\underset{k_n \to \infty}{\lim} \frac{1}{k_n}\sum\limits_{t=1}^{k_n} 1(y_{t-d}<a \text{ or } y_{t-d}>b)=0)=0$\\

$i.e., \underset{k_n \to \infty}{\lim}\prod\limits_{t=1}^{k_n} 1(a\leqslant y_{t-d}\leqslant b) \to 0$ as $k_n \to \infty$.\\

\noindent Therefore, we still have $\underset{\tiny{\bm{\theta \in \Theta}}}{\sup} \frac{1}{n}|\bm{\tilde{L}_n(\theta)-L_n(\theta)}|\to 0$. We finish the proof of $S1$.\\

\noindent Let us then show $S2$:\\

 $\begin{aligned}
 &E_{\theta_0}\{\bm{l_t(\theta)}\}-E_{\theta_0}\{\bm{l_t(\theta_0)}\}\\
 &=E_{\theta_0}\{[\bm{l_{1t}(\theta)-l_{1t}(\theta_0)}]R_{0t}R_t\} + E_{\theta_0}\{[\bm{l_{2t}(\theta)-l_{2t}(\theta_0)}](1-R_{0t})(1-R_t)\}\\
 &+E_{\theta_0}\{[\bm{l_{1t}(\theta)-l_{2t}(\theta_0)}](1-R_{0t})R_t\}+E_{\theta_0}\{[\bm{l_{2t}(\theta)-l_{1t}(\theta_0)}]R_{0t}(1-R_t)\}\\
 \end{aligned}$\\

 \quad\\
 $\begin{aligned}
 \text{Consider } &E_{\theta_0}[(\bm{l_{1t}(\theta)-l_{2t}(\theta_0))}R_{0t}R_t]\\
            &=E_{\theta_0}\{(\log\frac{\bm{\alpha_1^\prime X_{t-1}}}{\bm{\alpha_{10}^\prime X_{t-1}}}+\frac{\bm{\alpha_{10}^\prime X_{t-1}}}{\bm{\alpha_{1}^\prime X_{t-1}}}-1+\frac{[(\bm{\phi_{10}-\phi_1)^\prime Y_{t-1}}]^2}{\bm{\alpha_1^\prime X_{t-1}}})R_{0t}R_t\}\\
  \end{aligned}$\\

\noindent  By $\log\frac{1}{x}+x-1\geqslant 0$, we know:
  \begin{center}
   $E_{\theta_0}\{[\bm{l_{1t}(\theta)-l_{2t}(\theta)}]R_{0t}R_t\}\geqslant 0$ with equality iff $\frac{\bm{\alpha_1^\prime X_{t-1}}}{\bm{\alpha_{10}^\prime X_{t-1}}}=1$ and $(\bm{\phi_{10}-\phi_1)^\prime Y_{t-1}}=0, i.e., \bm{\alpha_1=\alpha_{10}}, \bm{\phi_1=\phi_{10}}$ \\
  \end{center}

\noindent  Similarly, we have:
  \begin{center}
  $E_{\theta_0}\{[\bm{l_{2t}(\theta)-l_{2t}(\theta_0)}](1-R_{0t})(1-R_t)\}\geqslant 0$ with equality iff $\frac{\bm{\alpha_2^\prime X_{t-1}}}{\bm{\alpha_{20}^\prime X_{t-1}}}=1$ and $(\bm{\phi_{20}-\phi_2)^\prime Y_{t-1}}=0, i.e., \bm{\alpha_2=\alpha_{20}}, \bm{\phi_2=\phi_{20}}$\\
  \end{center}

\noindent  For the third term, we have $E_{\theta_0}\{[\bm{l_{1t}(\theta)-l_{2t}(\theta_0)}](1-R_{0t})R_t\}\geq0$. Since we assume $\bm{\phi_{10}\neq \phi_{20}}$ and $\bm{\alpha_{10}\neq \alpha_{20}}$,  $E_{\theta_0}\{[\bm{l_{1t}(\theta)-l_{2t}(\theta_0)}](1-R_{0t})R_t\}=0$ implies $E_{\theta_0}\{(1-R_{0t})R_t\}=0$, from which it can be deduced, following the same argument in proving Theorem 2 in \cite*{li2015}, that $d=d_0, r_L\geqslant r_{0L}, r_U\geqslant r_{0U}$. Similarly, from the last term $E_{\theta_0}\{[\bm{l_{2t}(\theta)-l_{1t}(\theta_0)}]R_{0t}(1-R_t)\}=0$, we have $d=d_0, r_L \leqslant r_{0L}, r_U \leqslant r_{0U}$. We thus finish the proof of $S2$.\\

\noindent  Let us further show $S3$:\\

  $\bm{\theta}^*\in U_{\eta}(\bm{\theta})$\\
  \begin{equation}
  \label{S3}
  \begin{aligned}
  \bm{l_t(\theta^*)-l_t(\theta)}&=\{\bm{l_{1t}(\theta^*)-l_{1t}(\theta)}\}R_t(r_L^*, r_U^*, d)\\
                                     &+\bm{l_{1t}(\theta)}[R_t(r_L^*, r_U^*, d)-R_t(r_L, r_U, d)]\\
                                     &+[\bm{l_{2t}(\theta^*)-l_{2t}(\theta)}][1-R_t(r_L^*, r_U^*, d)]\\
                                     &+\bm{l_{2t}(\theta)}[R_t(r_L, r_U, d)-R_t(r_L^*, r_U^*, d)]\\
  \end{aligned}
  \end{equation}\\

  \begin{equation*}
  \begin{aligned}
  \bm{l_{1t}(\theta^*)-l_{1t}(\theta)}&=\log {\bm{\alpha_1^*}^\prime X_{t-1}}-\log \bm{\alpha_1^\prime X_{t-1}}\\
      &+\frac{(y_t-\bm{{\phi_1^*}^\prime Y_{t-1}})^2-(y_t-\bm{{\phi_1^\prime Y_{t-1}})^2}}{\bm{{\alpha_1^*}^\prime X_{t-1}}}\\
      &+(y_t-\bm{{\phi_1^\prime Y_{t-1}}})^2[\frac{1}{\bm{{\alpha_1^*}^\prime X_{t-1}}}-\frac{1}{\bm{\alpha_1^\prime X_{t-1}}}]
  \end{aligned}
  \end{equation*}\\

\noindent  By Taylor's expansion and a method similar to \cite{Zhu_Ling2013}, we have:\\

  $\begin{aligned}
  &E[\underset{\theta^*\in U_{\eta}(\theta)}{\sup} |\log {\bm{\alpha_1^*}^\prime X_{t-1}}-\log {\bm{\alpha_1}^\prime X_{t-1}}|]\to 0 \quad \text{as} \quad \eta \to 0\\
  &E[\underset{\theta^*\in U_{\eta}(\theta)}{\sup} |\frac{(y_t-\bm{{\phi_1^*}^\prime Y_{t-1}})^2-(y_t-\bm{{\phi_1}^\prime Y_{t-1}})^2}{\bm{{\alpha_1^*}^\prime X_{t-1}}}|]\to 0 \quad \text{as} \quad \eta \to 0\\
 &E[\underset{\theta^*\in U_{\eta}(\theta)}{\sup} |(y_t-\bm{{\phi_1^\prime Y_{t-1}}})^2(\frac{1}{\bm{{\alpha_1^*}^\prime X_{t-1}}}-\frac{1}{\bm{\alpha_1^\prime X_{t-1}}})|]\to 0 \quad \text{as} \quad \eta \to 0\\
  \end{aligned}$\\

\noindent  Therefore, the first and third terms in expression (\ref{S3}) can be dealt with. \\

\noindent  We now look at the second term in (\ref{S3}). \\

\noindent  Since $\bm{l_{1t}(\theta)}$ and $R_t(r_L^*, r_U^*,d)-R_t(r_L, r_U,d)$ have non-overlapping parameters,\\

  {\fontsize{9bp}{\baselineskip}$E\{\underset{\theta^* \in U_{\eta}(\theta)}{\sup} \bm{l_{1t}(\theta)}[R_t(r_L^*, r_U^*,d)-R_t(r_L, r_U,d)]\}=E\{\underset{\theta^* \in U_{\eta}(\theta)}{\sup} \bm{l_{1t}(\theta)} \underset{\theta^* \in U_{\eta}(\theta)}{\sup} [R_t(r_L^*, r_U^*,d)-R_t(r_L, r_U,d)]\}$}\\

\noindent  \cite*{li2015} shows that $E\{\underset{\theta^* \in U_{\eta}(\theta)}{\sup} |R_t(r_L^*, r_U^*,d)-R_t(r_L, r_U,d)|\}\to 0 \quad \text{as} \quad \eta \to 0$. Combining this with $E[\underset{\theta^* \in U_{\eta}(\theta)}{\sup} |\bm{l_{1t}(\theta)}|]<\infty$ and using dominated convergence theorem, we have:\\
  \begin{center}
   $E\{\underset{\theta^* \in U_{\eta}(\theta)}{\sup} \bm{l_{1t}(\theta)}[R_t(r_t^*, r_U^*, d)-R_t(r_t, r_U, d)]\}\to 0$ as $\eta \to 0$\\
  \end{center}

\noindent The fourth term in (\ref{S3}) can be dealt with in the same way. We thus finish the proof of $S3$.\\

\noindent Based on $S1$-$S3$ and \cite{Huber1967}, Theorem 2 is proved.\\
\subsection{Proof of Theorem 3}
 We first prove the super-consistency of the estimated boundary parameters at Claim (i).  Note in Theorem 2, we know $\bm{\hat{\theta}_n}$ is strong consistent. Therefore, without loss of generality, we restrict the parameter space to a neighborhood of $\bm{\theta_0}$.\\

  Define $w(\triangle)=\{\bm{\lambda} \in \bm{\land}, a<r_L<r_U<b: ||\bm{\lambda-\lambda_0}||<\triangle, |r_L-r_{0L}|<\triangle, |r_U-r_{0U}|<\triangle\},$ where $0<\triangle < \min\{1, \frac{r_{0U}-r_{0L}}{2}\}$ will be determined later. As in \cite{Chan1993}, it is sufficient to show for any $\epsilon>0, \exists$ a positive $K$ such that:\\
  \begin{equation}
  \label{Thm 3(i)}
  \text{Pr}(\tilde{L}_n(\bm{\lambda}, r_{0L}+z_L, r_{0U}+z_U)-\tilde{L}_n(\bm{\lambda}, r_{0L}, r_{0U})>0)>1-\epsilon
  \end{equation}

  where $\bm{\lambda}\in w(\triangle), |z_L|>\frac{K}{n}, |z_U|>\frac{K}{n}$.\\

  We consider the case $p=d=1$,$z_L <0$ and $z_U>0$. Denote the disjoint events:\\

  $\begin{aligned}
  & A_0=\{r_{0U}<y_{t-1} \leqslant r_{0U}+z_U, R_{t-1}=1\}\\
  & B_0=\{r_{0L}+z_L<y_{t-1} \leqslant r_{0L}, R_{t-1}=0 \}\\
  &A_i=\{y_{t-1}\in (r_{0L}, r_{0U}], \dots, y_{t-i}\in (r_{0L}, r_{0U}], r_{0U}<y_{t- i-1} \leqslant r_{0U}+z_U, R_{t-i-1}=1\}, i\geqslant 1\\
  &B_i=\{y_{t-1}\in (r_{0L}, r_{0U}], \dots, y_{t-i}\in (r_{0L}, r_{0U}], r_{0L}+z_L<y_{t- i -1} \leqslant r_{0L}, R_{t-i-1}=0 \}, i\geqslant 1\\
  \end{aligned}$\\

 where $R_t \equiv R_t(r_{0L}+z_L, r_{0U}+z_U)$. Denote $A_t(z_L, z_U)=\bigcup\limits_{j=0}^\infty A_j, B_t(z_L, z_U)=\bigcup\limits_{j=0}^\infty B_j$\\ As in \cite*{li2015}, we have:
 \begin{center}
 $R_t(r_{0L}+z_L, r_{0U}+z_U)-R_t(r_{0L}, r_{0U})=1\{A_t(z_L, z_U)\}-1\{B_t(z_L, z_U)\}$\\
  \end{center}

Moreover, $B_t(z_L, z_U) \subset \{R_t(r_{0L}, r_{0U})=1\}, A_t(z_L, z_U)\subset \{R_t(r_{0L}, r_{0U})=0\}$. As a result, \\

$\begin{aligned}
&\bm{L_n(\lambda}, r_{0L}+z_L, r_{0U}+z_U)-\bm{L_n(\lambda}, r_{0L}, r_{0U})\\
&=\sum\limits_{t=1}^n \bm{(l_{1t}(\theta)-l_{2t}(\theta))}[1\{A_t(z_L, z_U)\}-1\{B_t(z_L, z_U)\}]\\
&=\sum\limits_{t=1}^n \bm{(l_{1t}(\theta)-l_{2t}(\theta))}1\{A_t(z_L, z_U)\}+\sum\limits_{t=1}^n \bm{(l_{2t}(\theta)-l_{1t}(\theta))}1\{B_t(z_L, z_U)\}\\
&=\bm{L_{1n}}(z_L, z_U)+\bm{L_{2n}}(z_L, z_U)\\
  \end{aligned}$\\

  \quad \\
Our next step  is to show that

\begin{equation}\label{doexample}
\qquad\qquad \qquad\qquad  \text{Pr}(\bm{L_{1n}}(z_L, z_U)>0)>1-\epsilon
\end{equation}
when $\bm{\lambda}\in w(\triangle),-\triangle<z_L<0 , \frac{K}{n}<z_U<\triangle$.
\\
\\

$\begin{aligned}
\text{Define } &Q_{z_L}(z_U)=E[1\{A_t(z_L, z_U)\}]\\
                     &Q_{n, z_L}(z_U)=\frac{1}{n} \sum\limits_{t=1}^n 1\{A_t(z_L, z_U)\}\\
\end{aligned}$\\

By a calculation, we can obtain\\

$\begin{aligned}
&\frac{\bm{L_{1n}}(z_L, z_U)}{nQ_{z_L}(z_U)} \equiv \frac{\sum\limits_{t=1}^n [\bm{l_{1t}(\theta)-l_{2t}(\theta)}][1\{A_t(z_L, z_U)\}]}{nQ_{z_L}(z_U)}\\
&=\frac{\sum\limits_{t=1}^n [\log \frac{\bm{\alpha_{10}^\prime X_{t-1}}}{\bm{\alpha_{20}^\prime X_{t-1}}}+\frac{\bm{\alpha_{20}^\prime X_{t-1}}}{\bm{\alpha_{10}^\prime X_{t-1}}}-1+\frac{\{\bm{(\phi_{20}-\phi_{10})^\prime Y_{t-1}}\}^2}{\bm{\alpha_{10}^\prime X_{t-1}}}] 1\{A_t(z_L, z_U)\}}{nQ_{n, z_L}(z_U)}\\
&+\frac{\sum\limits_{t=1}^n \frac{2\bm{(\phi_{20}-\phi_{10})^\prime Y_{t-1}} \sqrt{\bm{\alpha_{20}^\prime X_{t-1}}}}{\bm{\alpha_{10}^\prime X_{t-1}}} \epsilon_t  1\{A_t(z_L, z_U)\}}{nQ_{n, z_L}(z_U)}\\
&+\frac{\sum\limits_{t=1}^n \frac{\bm{(\alpha_{20}-\alpha_{10})^\prime X_{t-1}}}{\bm{\alpha_{10}^\prime X_{t-1}}} (\epsilon_t^2-1)  1\{A_t(z_L, z_U)\}}{nQ_{n, z_L}(z_U)}+Op(\triangle)\\
\\
\end{aligned}$\\

Note in the first term,
\begin{center}
$\log \frac{\bm{\alpha_{10}^\prime X_{t-1}}}{\bm{\alpha_{20}^\prime X_{t-1}}}+\frac{\bm{\alpha_{20}^\prime X_{t-1}}}{\bm{\alpha_{10}^\prime X_{t-1}}}-1+\frac{\{\bm{(\phi_{20}-\phi_{10})^\prime Y_{t-1}}\}^2}{\bm{\alpha_{10}^\prime X_{t-1}}}>0$
\end{center}
where $\bm{X_{t-1}}=(1,y_{t-1}^2), \bm{Y_{t-1}}=(1, y_{t-1})$\\

In the second term, $\sum\limits_{t=1}^n \frac{2\bm{(\phi_{20}-\phi_{10})^\prime Y_{t-1}} \sqrt{\bm{\alpha_{20}^\prime X_{t-1}}}}{\bm{\alpha_{10}^\prime X_{t-1}}} \epsilon_t  1\{A_t(z_L, z_U)\}$ is bounded in absolute value by $K_1|\sum\limits_{t=1}^n \epsilon_t 1(A_t(z_L,z_U))|$\\

For the third term, the numerator is bounded in absolute value by $K_2|\sum\limits_{t=1}^n (\epsilon_t^2-1) 1(A_t(z_L,z_U))|$, where $K_1$ and $K_2$ are some constants independent of $n$.\\

Following the methond similar to Theorem 3 in \cite*{li2015},  we can verify the following three conditions:
$\forall \epsilon>0,\eta>0, \exists$ a positive constant $K$ such that as $n$ is large enough,\\
\begin{center}
$Pr[\underset{\frac{k}{n}<z_U<\triangle, -\triangle<z_L\leqslant 0}{\sup}|\frac{1}{nQ_{zL}(z_U)}\sum\limits_{t=1}^n 1\{A_t(z_L, z_U)\}-1|<\eta]>1-\epsilon$\\
$Pr[\underset{\frac{k}{n}<z_U<\triangle, -\triangle<z_L\leqslant 0}{\sup}|\frac{1}{nQ_{zL}(z_U)}\sum\limits_{t=1}^n \epsilon_t 1\{A_t(z_L, z_U)\}|<\eta]>1-\epsilon$\\
$Pr[\underset{\frac{k}{n}<z_U<\triangle, -\triangle<z_L\leqslant 0}{\sup}|\frac{1}{nQ_{zL}(z_U)}\sum\limits_{t=1}^n (\epsilon_t^2-1) 1\{A_t(z_L, z_U)\}|<\eta]>1-\epsilon$\\
\end{center}

Therefore, (\ref{doexample}) is implied. We can further show a similar result for $\bm{L_{2n}}(z_L, z_U)$, and then that for $L_n(\bm{\lambda}, r_{0L}+z_L, r_{0U}+z_U)-L_n(\bm{\lambda}, r_{0L}, r_{0U})$. (\ref{Thm 3(i)}) is thus proved for the case that $p=d=1$, $z_L<0$ and $z_U>0$.\\

 The proof for other cases is similar and hence is omitted. The proof of Claim (ii) in this theorem is similar to that of Theorem 2.2 in \cite*{Li2013} and that of Theorem 3.2 in \cite*{li2015b}, and is also omitted.\\

  \subsection{Proof of Theorem 4}
  From Theorem 3, we know $n(\hat{r}_L-r_{0L})=O_p(1)$ and $n(\hat{r}_U-r_{0U})=O_p(1)$. To further characterize their limiting distributions, we consider the limiting behavior of a sequence of normalized profile objective function defined by:
  \begin{center}
   $\bm{\tilde{L}_n}(z)=\bm{L_n}(\bm{\tilde{\lambda}}_n(r_0+\frac{z}{n}), r_0+\frac{z}{n})-\bm{L_n}(\bm{\tilde{\lambda}}_n(r_0), r_0)$\\
  \end{center}
  where $r_0=(r_{0L}, r_{0U})^T$ and $z=(z_L, z_U)^T$\\

  $\begin{aligned}
  \text{Denote } \varphi_n(z)&=\sum\limits_{t=1}\xi_{1t}\{1-R_{t-1}(r_0)\}1\{r_{0L}<y_{t-d}\leqslant r_{0L}+\frac{z_L}{n}\} 1(z_L\geqslant 0)\\
    &+\sum\limits_{t=1}\xi_{2t}\{1-R_{t-1}(r_0)\}1\{r_{0L}+\frac{z_L}{n}<y_{t-d}\leqslant r_{0L}\} 1(z_L< 0)\\
    &+\sum\limits_{t=1}\xi_{1t}R_{t-1}(r_0)1\{r_{0U}<y_{t-d}\leqslant r_{0U}+\frac{z_L}{n}\} 1(z_U\geqslant 0)\\
    &+\sum\limits_{t=1}\xi_{2t}R_{t-1}(r_0)1\{r_{0U}+\frac{z_U}{n}<y_{t-d}\leqslant r_{0U}\} 1(z_U< 0)\\
    \\
  \text{where  } &\xi_{1t}=\sum\limits_{j=0}^\infty \zeta_{1, t+j} H_{t+j, j}\\
                       &\xi_{2t}=\sum\limits_{j=0}^\infty \zeta_{2, t+j} H_{t+j, j}\\
                       &\zeta_{1t}=\log\frac{\bm{\alpha_{10}^\prime X_{t-1}}}{\bm{\alpha_{20}^\prime X_{t-1}}}+\frac{\{(\bm{\phi_{10}-\phi_{20}})^\prime \bm{Y_{t-1}}-\epsilon_t \sqrt{\bm{\alpha_{20}^\prime X_{t-1}}}\}^2}{\bm{\alpha_{10}^\prime X_{t-1}}}-\epsilon_t^2\\
                       &\zeta_{2t}=\log\frac{\bm{\alpha_{20}^\prime X_{t-1}}}{\bm{\alpha_{10}^\prime X_{t-1}}}+\frac{\{(\bm{\phi_{10}-\phi_{20}})^\prime \bm{Y_{t-1}}+\epsilon_t \sqrt{\bm{\alpha_{10}^\prime X_{t-1}}}\}^2}{\bm{\alpha_{20}^\prime X_{t-1}}}-\epsilon_t^2\\
                       &H_{t,j}=\prod\limits_{l=1}^j 1(r_{0L}<y_{t-d+1-l}\leqslant r_{0U})
  \end{aligned}$\\

 We first prove that, for any $B>0$, 
 \begin{equation}
   \label{proof}
   \qquad \qquad \underset{||z||\leqslant B}{\sup}|\bm{\tilde{L}_n}(z)-\varphi_n(z)|=o_p(1)\\
 \end{equation}\\
It is sufficient to verify the following two conditions:\\
\begin{equation}
   \label{straightforward}
   \underset{||z||\leqslant B}{\sup}|\bm{\tilde{L}_n}(z)-\{\bm{L_n}(\bm{\lambda_0}, r_0+\frac{z}{n})-\bm{L_n}(\bm{\lambda_0}, r_0)\}|=o_p(1)\\
 \end{equation}\\
 \begin{equation}
  \label{straightforward1}
    \underset{||z||\leqslant B}{\sup}|\varphi_n(z)-\{\bm{L_n}(\bm{\theta_0}, r_0+\frac{z}{n})-\bm{L_n}(\bm{\theta_0}, r_0)\}|=o_p(1)\\
 \end{equation}\\

By a method similar to similar to the proof of Theorem 4 in \cite*{li2015}, we can use Taylor's expansion and results of Theorem 3 in current paper to show that (\ref{straightforward}) holds. Below we will verify  (\ref{straightforward1}).  \\

  Denote $R_t(z)=R_t(r_0+\frac{z}{n})$ for simplicity. As in \cite*{li2015}, we have:

  \begin{equation*}
  \begin{aligned}
  R_t(z)-R_t(0)&=\{R_{t-1}(z)-R_{t-1}(0)\}1(r_{0L}<y_{t-d}\leqslant r_{0U})\\
   &+\{1-R_{t-1}(z)\}1(r_{0L}<y_{t-d}\leqslant r_{0L}+\frac{z_L}{n})1(z_L\geqslant 0)\\
   &-\{1-R_{t-1}(z)\}1(r_{0L}+\frac{z_L}{n}<y_{t-d}\leqslant r_{0L})1(z_L< 0)\\
   &+R_{t-1}(z)1(r_{0U}<y_{t-d}\leqslant r_{0U}+\frac{z_U}{n})1(z_U\geqslant 0)\\
   &-R_{t-1}(z)1(r_{0U}+\frac{z_U}{n}<y_{t-d}\leqslant r_{0U})1(z_U<0)\\
   \end{aligned}
  \end{equation*}

  \begin{equation}
  \begin{aligned}
  \label{R_t}
   \text{and then } R_t(z)-R_t(0)&=\sum\limits_{j=0}^\infty H_{t,j}\{1-R_{t-1-j}(z)\}1(r_{0L}<y_{t-d-j}\leqslant r_{0L}+\frac{z_L}{n}) 1(z_L\geqslant 0)\\
   &-\sum\limits_{j=0}^\infty H_{t,j}\{1-R_{t-1-j}(z)\}1(r_{0L}+\frac{z_L}{n}<y_{t-d-j}\leqslant r_{0L})1(z_L<0)\\
   &+\sum\limits_{j=0}^\infty H_{t,j}R_{t-1-j}(z)1(r_{0U}<y_{t-d-j}\leqslant r_{0U}+\frac{z_U}{n}) 1(z_U\geqslant 0)\\
   &-\sum\limits_{j=0}^\infty H_{t,j}R_{t-1-j}(z)1(r_{0U}+\frac{z_U}{n}<y_{t-d-j}\leqslant r_{0U}) 1(z_U<0)\\
   \end{aligned}
  \end{equation}
  where $H_{t,j}=\prod\limits_{l=1}^j 1(r_{0L}<y_{t-d+1-l} \leqslant r_{0U})$\\
\\

  $\begin{aligned}
  \text{Denote } \zeta_{t}&=\bm{l_{1t}(\theta_0)-l_{2t}(\theta_0)}\\
                                &=\log\bm{\alpha_{10}^\prime X_{t-1}}+\frac{(y_t-\bm{\phi_{10}^\prime Y_{t-1}})^2}{\bm{\alpha_{10}^\prime X_{t-1}}}-\log\bm{\alpha_{20}^\prime X_{t-1}}-\frac{(y_t-\bm{\phi_{20}^\prime Y_{t-1}})^2}{\bm{\alpha_{20}^\prime X_{t-1}}}\\
  \end{aligned}$\\

     and it holds that $\bm{L_n}(\bm{\lambda_0},r_0+\frac{z}{n})-\bm{L_n}(\bm{\lambda_0},r_0)=\sum\limits_{t=1}^n \zeta_t\{R_t(z)-R_t(0)\}$.\\

     For the first term of (\ref{R_t}), i.e., the case with $z_L\geqslant 0$, by a method similar to \cite*{li2015}, it can be shown that:
     \begin{center}
     $\underset{||z||\leqslant B}{\sup}||\sum\limits_{t=1}^n \zeta_t \sum\limits_{j=0}^\infty H_{t,j}\{R_{t-1-j}(z)-R_{t-1-j}(0)\}1(r_{0L}<y_{t-d-j}\leqslant r_{0L}+\frac{z_L}{n})||=o_p(1)$\\
     \end{center}
\space{\ }

Note $\zeta_t=\zeta_{1t}$ when $R_t(0)=0$ and $\zeta_t=-\zeta_{2t}$ when $R_t(0)=1$. Thus, we have\\

    $\begin{aligned}
    &\sum\limits_{t=1}^n \zeta_t \sum\limits_{j=0}^\infty H_{t,j}\{1-R_{t-1-j}(0)\} 1(r_{0L}<y_{t-d-j} \leqslant r_{0L}+\frac{z_L}{n})\\
    &=\sum\limits_{j=0}^\infty \sum\limits_{t=1-j}^{n-j} \zeta_{t+j} H_{t+j,j} \{1-R_{t-1}(0)\} 1(r_{0L}<y_{t-d} \leqslant r_{0L}+\frac{z_L}{n})\\
    &=\sum\limits_{j=0}^\infty (\sum\limits_{t=1}^n+\sum\limits_{t=1-j}^0-\sum\limits_{t=n-j+1}^n)\zeta_{t+j} H_{t+j,j} \{1-R_{t-1}(0)\} 1(r_{0L}<y_{t-d} \leqslant r_{0L}+\frac{z_L}{n})\\
    &=\sum\limits_{j=0}^\infty \sum\limits_{t=1-j}^{n} \zeta_{t+j} H_{t+j,j} \{1-R_{t-1}(0)\} 1(r_{0L}<y_{t-d} \leqslant r_{0L}+\frac{z_L}{n})+o_p(1)\\
    &=\sum\limits_{t=1-j}^{n} \xi_{1t} \{1-R_{t-1}(0)\} 1(r_{0L}<y_{t-d} \leqslant r_{0L}+\frac{z_L}{n})+o_p(1)\\
    \end{aligned}$\\
 where the $o_p(1)$ is uniformly for $||z||\leqslant B$. See also the proof of Theorem 4 in \cite*{li2015} and Lemma 8.2 in \cite*{Li2013}.\\

 Similarly, we can handle the remaining terms in (\ref{R_t}). Therefore,  (\ref{straightforward1}) is implied.  With (\ref{straightforward}) and  (\ref{straightforward1}), it is clear that (\ref{proof}) holds.\\

 For the space $\mathbb{D(R}^2)$, we define the skorohod metric as $d(f,g)=\sum\limits_{k=1}^\infty 2^{-k}\min\{1, d_k(f,g)\}$ for $f,g \in \mathbb{D(R}^2)$, where $d_k(f,g)$ is the skorohod metric on $\mathbb{D}([-k,k]\times[-k,k])$, see also \cite*{li2015}, \cite{Li2012} and (16.4) in \cite{Billingsley}.\\

    By a technique similar that used in the proof of Theorem 3.3 in \cite*{Li2012} and in the proof of Theorem 5 in \cite*{li_ling2015a}, together with Theorem 5.5 in \cite{Straf1972}, we are able to conclude that $\{\varphi_n(z), z\in \mathbb{R}^2\}$ converges weakly to $\{\varphi(z), z\in \mathbb{R}^2\}$ as $n\to \infty$.\\

    Moreover, (\ref{proof}) implies that $d(\tilde{L}_n(z),\varphi_n(z)) \to 0$ in probability as $n \to \infty$. By Theorem 3.1 in \cite{Seijo2011}, it is readily seen that $n(\hat{r}_L-r_{0L})\to M_{-}^{(L)}$ and $n(\hat{r}_U-r_{0U})\to M_{-}^{(U)}$ in distribution as $n\to \infty$ respectively. The remainder of the proof is similar to that of Theorem 2 in \cite{Chan1993}).\\
\subsection{Proof of Theorem 5}
 From Theorem 2 and 3, $\hat{r}_L$ and $\hat{r}_U$ are super-consistent, and $\hat{d}$ is consistent with integer values. Therefore, it can be assumed that the true values of $(r_L, r_U, d)$ are known, indicating the true Regime indicators $R_t$ are known.\\

\noindent By definition, \\
$\begin{aligned}
\text{BIC}(p)&=\sum\limits_{t=1}^n \bm{l_t(\hat{\theta})}+(2p+2)\log n_1+(2p+2)\log n_2\\
    &=\sum\limits_{t=1}^n \bm{l_{1t}(\hat{\theta})}R_t+\sum\limits_{t=1}^n \bm{l_{2t}(\hat{\theta})}(1-R_t)+(2p+2)\log n_1+(2p+2)\log n_2\\
    &=\sum\limits_{t=1}^n \bm{l_{1t}(\hat{\lambda}_1)}R_t+\sum\limits_{t=1}^n \bm{l_{2t}(\hat{\lambda}_2)}(1-R_t)+(2p+2)\log n_1+(2p+2)\log n_2\\
\end{aligned}$\\
where $\bm{l_{1t}(\theta)}=\bm{l_{1t}(\lambda_1)}$ and $\bm{l_{2t}(\theta)}=\bm{l_{2t}(\lambda_2)}$ are defined at the beginning of the proof of Theorem 2. \\

    We first consider the case $p>p_0$. Below we introduce notations $\bm{\lambda}^p$,$\bm{\lambda}^p_0$ and $\hat{\bm{\lambda}}^p$ to emphasize their dependence on the order p. Under the situation $p>p_0$, the model with order $p$ corresponds to a bigger model, and we have
    \begin{center}
     $\bm{l_{1t}}(\bm{\lambda}_{10}^p)=\bm{l_{1t}}(\bm{\lambda}_{10}^{p_0}), \bm{l_{2t}}(\bm{\lambda}_{20}^p)=\bm{l_{2t}}(\bm{\lambda}_{20}^{p_0})$\\
    \end{center}

    By an intermediate result in the proof of Theorem 4 in \cite{Li2016}, we know\\

    $\begin{aligned}
   \qquad \qquad   &\sum\limits_{t=1}^n \bm{l_{1t}}(\bm{\hat{\lambda}}_1^{p_0})R_t-\sum\limits_{t=1}^n \bm{l_{1t}}(\bm{\lambda}_{10}^{p_0})R_t=O_p(1)\\
       &\sum\limits_{t=1}^n \bm{l_{1t}}(\bm{\hat{\lambda}}_1^{p})R_t-\sum\limits_{t=1}^n \bm{l_{1t}}(\bm{\lambda}_{10}^{p})R_t=O_p(1)\\
       &\sum\limits_{t=1}^n \bm{l_{2t}}(\bm{\hat{\lambda}}_2^{p_0})(1-R_t)-\sum\limits_{t=1}^n \bm{l_{2t}}(\bm{\lambda}_{20}^{p_0})(1-R_t)=O_p(1)\\
       &\sum\limits_{t=1}^n \bm{l_{2t}}(\bm{\hat{\lambda}}_2^{p})(1-R_t)-\sum\limits_{t=1}^n \bm{l_{2t}}(\bm{\lambda}_{20}^{p})(1-R_t)=O_p(1)\\
\end{aligned}$
       \\
As a result,\\
    $\begin{aligned}
    \sum\limits_{t=1}^n \bm{l_{1t}}(\bm{\hat{\lambda}}_1^{p})R_t-\sum\limits_{t=1}^n \bm{l_{1t}}(\bm{\hat{\lambda}}_{1}^{p_0})R_t=&[\sum\limits_{t=1}^n \bm{l_{1t}}(\bm{\hat{\lambda}}_1^{p})R_t-\sum\limits_{t=1}^n \bm{l_{1t}}(\bm{\lambda}_{10}^{p})R_t]\\
&-[\sum\limits_{t=1}^n \bm{l_{1t}}(\bm{\hat{\lambda}}_1^{p_0})R_t-\sum\limits_{t=1}^n \bm{l_{1t}}(\bm{\lambda}_{10}^{p_0})R_t]\\
&+[\sum\limits_{t=1}^n \bm{l_{1t}}(\bm{\lambda}_{10}^{p})R_t]-\sum\limits_{t=1}^n \bm{l_{1t}}(\bm{\lambda}_{10}^{p_0})R_t]=O_p(1)\\
\end{aligned}$\\
    $\begin{aligned}
\sum\limits_{t=1}^n \bm{l_{2t}}(\bm{\hat{\lambda}}_2^{p})(1-R_t)-\sum\limits_{t=1}^n \bm{l_{2t}}(\bm{\hat{\lambda}}_{2}^{p_0})(1-R_t)=&[\sum\limits_{t=1}^n \bm{l_{2t}}(\bm{\hat{\lambda}}_2^{p})(1-R_t)-\sum\limits_{t=1}^n \bm{l_{2t}}(\bm{\lambda}_{20}^{p})(1-R_t)]\\
&-[\sum\limits_{t=1}^n \bm{l_{2t}}(\bm{\hat{\lambda}}_2^{p_0})(1-R_t)-\sum\limits_{t=1}^n \bm{l_{2t}}(\bm{\lambda}_{20}^{p_0})(1-R_t)]\\
&+[\sum\limits_{t=1}^n \bm{l_{2t}}(\bm{\lambda}_{20}^{p})(1-R_t)-\sum\limits_{t=1}^n \bm{l_{2t}}(\bm{\lambda}_{20}^{p_0})(1-R_t)]\\
&=O_p(1)\\
    \end{aligned}$\\

Therefore, $\text{BIC}(p)-\text{BIC}(p_0)=O_p(1)+2(p-p_0)\log n_1n_2 \to \infty$ as $n\to \infty$, since $\frac{n_1}{n}\to \text{Pr}(R_t=1)>0$ and $\frac{n_2}{n}\to \text{Pr}(R_t=0)>0$ as $n\to \infty$.\\

    We then consider the case $p<p_0$. Again by an intermediate result in the proof of Theorem 4 in \cite{Li2016}, we know
    \begin{center}
       $\sum\limits_{t=1}^n \bm{l_{1t}}(\bm{\hat{\lambda}}_1^{p})R_t-\sum\limits_{t=1}^n \bm{l_{1t}}(\bm{\hat{\lambda}_{1}^{p_0}})R_t=O_p(1)+o_p(n_1)+c_1n_1$\\
       $\sum\limits_{t=1}^n \bm{l_{2t}}(\bm{\hat{\lambda}}_2^{p})(1-R_t)-\sum\limits_{t=1}^n \bm{l_{2t}}(\bm{\hat{\lambda}}_{2}^{p_0})(1-R_t)=O_p(1)+o_p(n_2)+c_2n_2$\\
    \end{center}
    where $c_1, c_2$ are positive constants, defined in the same way as constant c in \cite{Li2016}.\\

    $\begin{aligned}
    \text{Therefore, } \text{BIC}(p)-\text{BIC}(p_0)&=c_1n_1+c_2n_2+O_p(1)+o_p(n_1)+o_p(n_2)+2(p-p_0)\ln n_1n_2\\
        &=c_3n+O_p(1)+o_p(n)+O(\ln n)\\
        &\to \infty \text{ as } n\to \infty\\
    \end{aligned}$\\
    since constant $c_3>0$, $\frac{n_1}{n}\to \text{Pr}(R_t=1)>0$ and $\frac{n_2}{n}\to \text{Pr}(R_t=0)>0$ as $n\to \infty$.\\


\newpage

\begin{sidewaystable}
\begin{center}
\caption{\label{table1}Simulation results for buffered double autoregressive model}\vspace{5mm}
\footnotesize
\begin{tabular}{rlrrrrrrcrrrrrrrcrrr}
\hline\hline
&&\multicolumn{6}{c}{1st regime}&&\multicolumn{6}{c}{2nd regime}&&\\
\cline{3-8}\cline{10-15}
$n$&&$\phi_{10}$&$\phi_{11}$&$\phi_{12}$&$\alpha_{10}$&$\alpha_{11}$&$\alpha_{12}$&&
     $\phi_{20}$&$\phi_{21}$&$\phi_{22}$&$\alpha_{20}$&$\alpha_{21}$&$\alpha_{22}$&&$r_L$&$r_U$\\
\hline
 400&Bias&0.0020& -0.0013&-0.0124&-0.0024&-0.0102&0.0054&&-0.0039&-0.0008&-0.0041&-0.0010&0.0010&-0.0045&&0.0013&-0.0102\\
    &ESD&0.0288& 0.0883&0.0765&0.0165&0.1161&0.0657&&0.0225&0.0835&0.0680&0.0103&0.0959 &0.0693&&0.0421&0.0314\\
    &ASD&0.0255& 0.0798&0.0723&0.0152&0.1057&0.0719&&0.0197&0.0681&0.0627&0.0086&0.0741&0.0571&&&\\

800&Bias& 0.0028& -0.0025&-0.0030&-0.0005&-0.0037&-0.0030&&-0.0004&-0.0055&-0.0020&-0.0005&0.0007&-0.0039&&0.0005&-0.0039\\
    &ESD & 0.0188& 0.0616& 0.0529& 0.0112&0.0855&0.0471&&0.0155&0.0503&0.0467&0.0068&0.0631&0.0494&&0.0173&0.0165\\
    &ASD& 0.0184& 0.0573&0.0518&0.0113&0.0786&0.0506&&0.0142&0.0486&0.0448&0.0063&0.0551&0.0414
&&&\\

\hline
\end{tabular}
\end{center}
\end{sidewaystable}

\newpage

\begin{figure}
\centering \scalebox{0.6}[0.8]{
\includegraphics{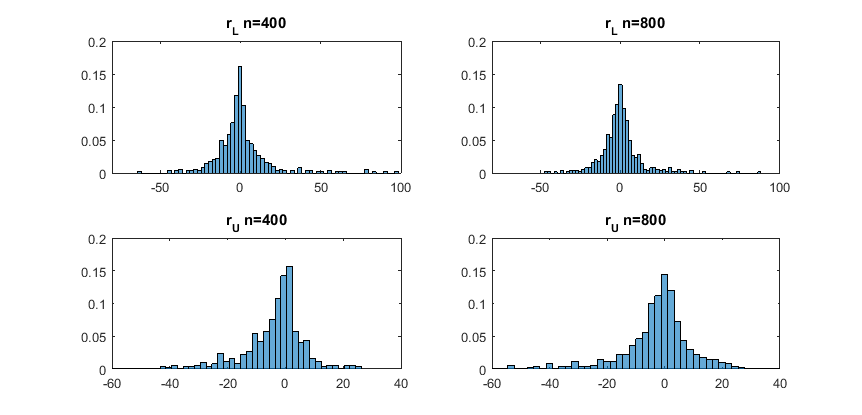}}
\caption{\label{fig1} Histograms of random variables $n(\hat{r}_L-r_L)$(upper panels) and $n(\hat{r}_U-r_U)$(lower panels)}
\end{figure}

\begin{figure}
\centering \scalebox{0.6}[0.6]{
\includegraphics{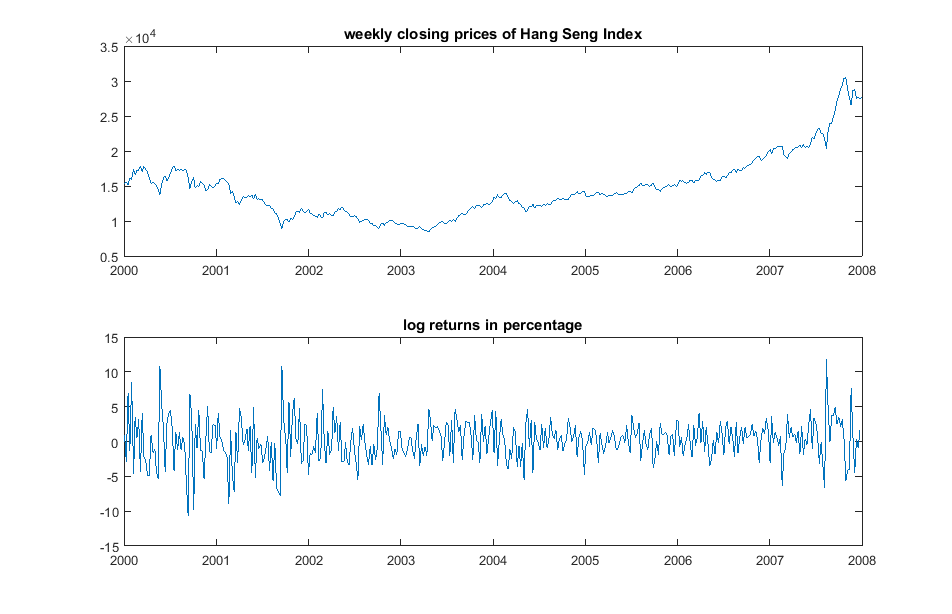}}
\caption{\label{fig2} Time plots of weekly closing prices of Hang Seng Index (HSI) from January 2000 to December 2007(upper panel) and corresponding log retuns(lower panel)}
\end{figure}

\begin{figure}
\centering \scalebox{0.8}[0.8]{
\includegraphics{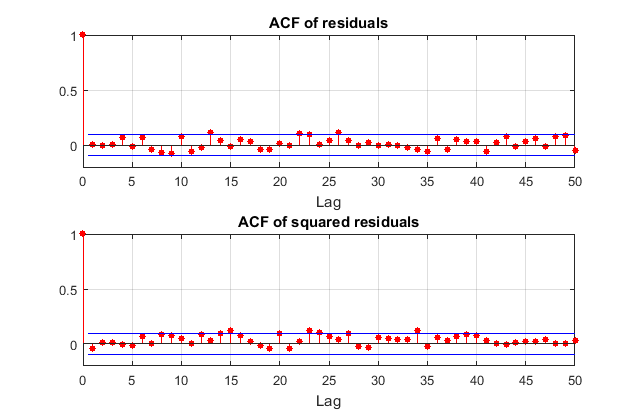}}
\caption{\label{fig3} Sample ACFs of residuals and squared residuals of fitted buffered double autoregressive model in empirical analysis}
\end{figure}


\newpage

\begin{thebibliography}{}

\bibitem[\protect\citeauthoryear{Billingsley}{Billingsley}{1999}]{Billingsley}
Billingsley, P. (1999).
\newblock {\em Convergence of Probability Measures}.
\newblock New York: Wiley, 2nd ed.

\bibitem[\protect\citeauthoryear{Chan}{Chan}{1993}]{Chan1993}
Chan, K.~S. (1993).
\newblock Consistency and limiting distribution of the least squares estimator
  of a threshold autoregressive model.
\newblock {\em The Annals of Statistics\/}~{\em 21}, 520--533.

\bibitem[\protect\citeauthoryear{Huber}{Huber}{1967}]{Huber1967}
Huber, P.~J. (1967).
\newblock The behavior of maximum likelihood estimates under nonstandard
  conditions.
\newblock In {\em Proceedings of the Fifth Berkeley Symposium on Mathematical
  Statistics and Probability}, Volume~1, Berkeley, pp.\  221--233. University
  of California Press.

\bibitem[\protect\citeauthoryear{Lee}{Lee}{2006}]{Lee2006}
Lee, O. (2006).
\newblock Stationarity and $\beta$-mixing property of a mixture ar-arch models.
\newblock {\em Bull. Korean Math. Soc\/}~{\em 43}, 813--820.

\bibitem[\protect\citeauthoryear{Li and Ling}{Li and Ling}{2012}]{Li2012}
Li, D. and S.~Ling (2012).
\newblock On the least squares estimation of multiple-regime threshold
  autoregressive models.
\newblock {\em Journal of Econometrics\/}~{\em 167}, 240--253.

\bibitem[\protect\citeauthoryear{Li, Ling, and Li}{Li et~al.}{2013}]{Li2013}
Li, D., S.~Ling, and K.~Li, W (2013).
\newblock Asymptotic theory on the least squares estimation of threshold
  moving-average model.
\newblock {\em Econometric Theory\/}~{\em 29}, 482--516.

\bibitem[\protect\citeauthoryear{Li, Ling, and Zakoian}{Li
  et~al.}{2015}]{li2015b}
Li, D., S.~Ling, and J.~Zakoian (2015).
\newblock Asymptotic inference in multiple-threshold double autoregressive
  models.
\newblock {\em Journal of Econometrics\/}~{\em 189}, 415--427.

\bibitem[\protect\citeauthoryear{Li, Ling, and Zhang}{Li
  et~al.}{2016}]{li_ling2015a}
Li, D., S.~Ling, and R.~Zhang (2016).
\newblock On a threshold double autoregressive model.
\newblock {\em Journal of Business and Economic Statistics\/}~{\em 34}, 68--80.

\bibitem[\protect\citeauthoryear{Li, Guan, Li, and Yu}{Li
  et~al.}{2015}]{li2015}
Li, G., B.~Guan, W.~K. Li, and P.~L.~H. Yu (2015).
\newblock Hysteretic autoregressive time series models.
\newblock {\em Biometrika\/}~{\em 102}, 717--723.

\bibitem[\protect\citeauthoryear{Li, Zhu, Liu, and Li}{Li
  et~al.}{2017}]{Li2016}
Li, G., Q.~Zhu, Z.~Liu, and W.~K. Li (2017).
\newblock On mixture double autoregressive time series models.
\newblock {\em Journal of Business and Economic Statistics\/}~{\em 35},
  306--317.

\bibitem[\protect\citeauthoryear{Ling}{Ling}{2004}]{ling2004}
Ling, S. (2004).
\newblock Estimation and testing stationarity for double-autoregressive models.
\newblock {\em Journal of the Royal Statistical Society, Series B\/}~{\em 66},
  63--78.

\bibitem[\protect\citeauthoryear{Ling}{Ling}{2007}]{ling2007}
Ling, S. (2007).
\newblock A double \textsc{ar}(p) model: structure and estimation.
\newblock {\em Statistica Sinica\/}~{\em 17}, 161--175.

\bibitem[\protect\citeauthoryear{Lo, Li, and Li}{Lo et~al.}{2016}]{lo2015}
Lo, P., W.~Li, and G.~Li (2016).
\newblock On buffered threshold garch model.
\newblock {\em Statistica Sinica\/}~{\em 26}, 1555--1567.

\bibitem[\protect\citeauthoryear{Lu}{Lu}{1998}]{Lu1998}
Lu, Z. (1998).
\newblock On the geometric ergodicity of a non-linear autoregressive model with
  an autoregressive conditional heteroscedastic term.
\newblock {\em Statistica Sinica\/}~{\em 8}, 1205--1217.

\bibitem[\protect\citeauthoryear{Mcleod and Li}{Mcleod and
  Li}{1983}]{Mcleod1983}
Mcleod, A. and W.~K. Li (1983).
\newblock Diagnostic checking arma time series models using squared residual
  autocorrelations.
\newblock {\em Journal of Time Series Analysis\/}~{\em 4}, 269--273.

\bibitem[\protect\citeauthoryear{Seijo and Sen}{Seijo and
  Sen}{2011}]{Seijo2011}
Seijo, E. and B.~Sen (2011).
\newblock A continuous mapping theorem for the smallest argmax functional.
\newblock {\em Electronic Journal of Statistics\/}~{\em 5}, 421--439.

\bibitem[\protect\citeauthoryear{Straf}{Straf}{1972}]{Straf1972}
Straf, M.~L. (1972).
\newblock Weak convergence of stochastic processes with several parameters.
\newblock {\em In Proceedings of the Sixth Berkeley Symposium on Mathematical
  Statistics and Probability, Vol. II: Probability Theory, Berkeley,
  California\/}.

\bibitem[\protect\citeauthoryear{Tong}{Tong}{1990}]{Tong1990}
Tong, H. (1990).
\newblock {\em Nonlinear Time Series: A Dynamical System Approach}.
\newblock Oxford: Oxford University Press.

\bibitem[\protect\citeauthoryear{Tsay}{Tsay}{1986}]{Tsay1986}
Tsay, R.~S. (1986).
\newblock Nonlinearility test for time series.
\newblock {\em Biometrika\/}~{\em 73}, 461--466.

\bibitem[\protect\citeauthoryear{Tse}{Tse}{2002}]{Tse2002}
Tse, Y.~K. (2002).
\newblock Residual-based diagnostics for conditional heteroscedasticity models.
\newblock {\em The Econometrics Journal\/}~{\em 5}, 358--373.

\bibitem[\protect\citeauthoryear{Tweedie}{Tweedie}{1983}]{Tweedie1983}
Tweedie, R.~L. (1983).
\newblock Criteria for rates of convergence of markov chains, with application
  to queueing and storage theory.
\newblock In J.~F.~C. Kingman and G.~E.~H. Reuter (Eds.), {\em Probability,
  Statistics and Analysis}. Cambridge: Cambridge University Press.

\bibitem[\protect\citeauthoryear{Zhu and Ling}{Zhu and
  Ling}{2013}]{Zhu_Ling2013}
Zhu, K. and S.~Ling (2013).
\newblock Quasi-maximum exponential likelihood estimators for a double
  \textsc{ar}(p) model.
\newblock {\em Statistica Sinica\/}~{\em 23}, 251--270.

\bibitem[\protect\citeauthoryear{Zhu, Yu, and Li}{Zhu et~al.}{2014}]{zhu2014}
Zhu, K., P.~L.~H. Yu, and W.~K. Li (2014).
\newblock Testing for the buffered autoregressive processes.
\newblock {\em Statistica Sinica\/}~{\em 24}, 971--984.

\bibitem[\protect\citeauthoryear{Zhu, Yu, and Li}{Zhu et~al.}{2017}]{zhu2015}
Zhu, K., P.~L.~H. Yu, and W.~K. Li (2017).
\newblock Buffered autorgressive models with conditional heteroscedasticity: An
  application to exchange ratess.
\newblock {\em Journal of Businees and Economic Statistics\/}~{\em 35},
  528--542.

\end{thebibliography}
\end{document}